\documentclass[twocolumn,aps,prl,superscriptaddress]{revtex4-1}

\usepackage{slashed}
\usepackage{tikz-cd}
\usepackage{tikz}
\usepackage{amsmath}
\usepackage{amsfonts}
\usepackage{bbold}
\usepackage{amssymb}
\usepackage{graphicx}
\usepackage{mathtools}
\usepackage{stmaryrd}
\usepackage{bbm}	
\usepackage{tikz}
\usetikzlibrary{decorations.pathreplacing,calligraphy,decorations.markings}

\usepackage[colorlinks=true,citecolor=blue,linkcolor=red]{hyperref}

\definecolor{tensor}{rgb}{0.5,0.8,0.5}
\definecolor{isometry}{rgb}{0.8,0.8,1}
\definecolor{unitary}{rgb}{0.8,0.8,1}
\definecolor{gate}{rgb}{1.0,1.0,1.0}

\newcommand{\ATensorLPDO}[2]{
	\begin{scope}[shift={(#1)}]
		\draw (-1.6,0) -- (1.6,0);
		\draw(-0.6,1.2) -- (-0.6,0);
		\draw[red] (0.6,1.2) -- (0.6,0);
		\filldraw[fill=tensor] (-0.9,-0.6) -- (-0.9,0.6) -- (0.9,0.6) -- (0.9,-0.6) -- (-0.9,-0.6);
		\draw (0,0) node {\scriptsize #2};
	\end{scope}
}

\newcommand{\ADaggTensorLPDO}[2]{
	\begin{scope}[shift={(#1)}]
		\draw (-1.6,0) -- (1.6,0);
		\draw (-0.6,-1.2) -- (-0.6,0);
		\draw[red] (0.6,-1.2) -- (0.6,0);
		\filldraw[fill=tensor] (-0.9,-0.6) -- (-0.9,0.6) -- (0.9,0.6) -- (0.9,-0.6) -- (-0.9,-0.6);
		\draw (0,0) node {\scriptsize #2};
	\end{scope}
}

\newcommand{\uTensor}[2]{
	\begin{scope}[shift={(#1)}]
		\draw (0,-0.3) --(0,1.3);	   \filldraw[fill=unitary,shift={(0,0)}](0,0.5) circle (0.5);
		\draw (0,0.5) node {\scriptsize #2};
	\end{scope}
}

\newcommand{\uTensorA}[2]{
	\begin{scope}[shift={(#1)}]
		\draw[red] (0,-0.3) --(0,1.3);	   \filldraw[fill=unitary,shift={(0,0)}](0,0.5) circle (0.5);
		\draw (0,0.5) node {\scriptsize #2};
	\end{scope}
 }

\newcommand{\vTensor}[2]{
	\begin{scope}[shift={(#1)}]
		\draw (1.3,0) --(-0.3,0);
        \filldraw[fill=unitary,shift={(0,0)}](0.5,0) circle (0.5);
		\draw (0.5,0) node {\scriptsize #2};
	\end{scope}
}

\definecolor{ZM}{rgb}{0,0,1} 
\definecolor{XQ}{rgb}{1,0,0}
\definecolor{XQ}{rgb}{0,0,0} 
\definecolor{SD}{rgb}{1.5,0,1.5}

\begin{document}
\title{Topological response in open quantum systems with weak symmetries}
\author{Ze-Min Huang}
\affiliation{Institute for Theoretical Physics, University of Cologne, 50937 Cologne, Germany}

\author{Sebastian Diehl}
\affiliation{Institute for Theoretical Physics, University of Cologne, 50937 Cologne, Germany}

\author{Xiao-Qi Sun}\email{xiaoqi.sun@mpq.mpg.de}
\affiliation{Max Planck Institute of Quantum Optics, Hans-Kopfermann-Stra{\ss}e 1, D-85748 Garching, Germany}
\affiliation{Munich Center for Quantum Science and Technology (MCQST), Schellingstra{\ss}e 4, D-80799 M{\"u}nchen, Germany}
\date{\today}

\begin{abstract}
In open quantum systems, the interaction of the system with its environment gives rise to two types of symmetry: a strong one, where the system's symmetry charge is conserved exactly, and a weak one, where the system can exchange symmetry charge with the environment but still preserve symmetry at the ensemble level. 
While generic open quantum systems feature weak symmetries only, the symmetry protected topological response for bosonic/spin systems has only been considered in the stricter setup with additional strong symmetries. Here, we address the generic case and demonstrate that weak symmetries alone can protect topological responses that distinguish different phases of matter. For bosonic systems, focusing on one-dimensional mixed states described by locally purifiable density operators, we propose a quantized response characterizing qualitatively distinct phases. It is detectable via the decay behavior of different string order parameters. We illustrate our general results through a noisy Affleck-Kennedy-Lieb-Tasaki model. In particular, we show that the coupling to the environment can induce a phase transition to a state protected by weak symmetries, without a pure-state or strong-symmetry analog.
\end{abstract}

\maketitle 
\textcolor{red}{\emph{Introduction.--}} Inevitable coupling to the environment transforms pure states into mixed-state ensembles, enriching the interplay between symmetry and topology.
In particular, the notion of symmetry is refined into two classes,  \emph{weak} and \emph{strong} symmetries~\cite{buca2012njp}, distinguished by whether the symmetry charge can be exchanged with the environment, or not.  Considerable efforts have been made to pursue these new aspects, broadly divided into two categories. For fermionic topological phases \cite{viyuela2014prl1d,budich2015prb, bardyn2018prx, huang2022prb,altland2021prx,huang2024arxiv_mstop, mao2024rpp, huang2025prl} with weak symmetry~\cite{altland2021prx}, topological order parameters have been constructed in all symmetry classes~\cite{huang2024arxiv_mstop, huang2025prl}. 
Meanwhile, for bosonic (spin) systems, nontrivial symmetry-protected topological (SPT) phases have been established only under strict conditions that require the presence of at least one \emph{strong symmetry}~\cite{ma2023prx,coser2019quantum,degroot2022quantum,guo2024arxiv_lpdo,xue2024arxiv_mpdo,kawabata2024prl,lee2025quantum}. However, generic  open quantum systems exhibit only \emph{weak symmetries}, making topological signatures without fine-tuning in these bosonic systems challenging at first sight.
A robust topological signature protected solely by weak symmetries is thus highly desirable, and could be detected on emerging experimental platforms, e.g., trapped ions \cite{blatt2012nature,monroe2021rmp,noel2022np}, Rydberg atoms \cite{semeghini2021science}, and superconducting circuits \cite{satzinger2021science}, which offer capabilities beyond traditional solid-state setups~\cite{blatt2012nature,monroe2021rmp,noel2022np,gross2017science, semeghini2021science, ebadi2021nature, satzinger2021science,iqbal2024nature, acharya2024nature,chen2025np}. Specifically, these platforms enable the measurement of non-local observables, e.g., full counting statistics in optical lattices~\cite{gross2017science} and string order parameters in programmable quantum simulators~\cite{semeghini2021science, iqbal2024nature}.

In this letter, we consider the more physical setup for bosonic systems with only weak symmetries. Focusing on one dimension, we rigorously demonstrate robust topological properties, characterized by quantized symmetry charge responses to twisted boundary conditions (or symmetry flux insertion). In particular, this quantized response defines topological invariants that extend beyond current frameworks, and are experimentally detectable via qualitatively distinct behaviors of string order parameters across different phases. This exact result is based on two ingredients: (i) a finite on-site weak symmetry group, where the quantized response is  defined given a pair of commuting elements; and (ii) a tensor network ansatz~\cite{verstrate2004prl, zwolak2004prl, cuevas2013njp}, where mixed states are efficiently described by locally purified density operators (LPDOs) and the purifications are assumed to be short-range correlated. We illustrate these results in an exemplary mixed state resulting from the Affleck-Kennedy-Lieb-Tasaki (AKLT) state exposed to on-site weak-symmetry preserving noise channels. Remarkably, we observe a noise-driven transition to  a mixed-state phase with quantized responses that cannot occur in previously studied systems with strong symmetries.

\textcolor{red}{\emph{Quantized response in pure states.--}} To set the stage for mixed states, we first recap relevant concepts of quantized response in pure SPT states~\cite{pollmann2010prb,chen2011prb, schuch2011prb,senthil2015arcmp} in the language of matrix product states (MPSs)~\cite{zaletel2013prl,zaletel2014njp,zaletel2014prb,cirac2021rmp}. Consider a topological state $|\psi\rangle$ of a one-dimensional spin chain protected by an on-site~\footnote{Here the requirement of a well-defined linear on-site representation is to ensure the stability of SPTs~\cite{pollmann2017oxford,anfuso2007fragility}.} finite-group symmetry $G$, i.e., $U_g|\psi\rangle\sim|\psi\rangle$ up to a phase factor. 
Here, $g\in G$ labels a group element and $U_g=u_g^{\otimes N}$ with $N$ being the system size. 
Recall the following MPS representation with finite bond dimension:
\begin{equation}
|\psi(X)\rangle =\sum_{\{i_j\}}\text{tr}(X\ A^{i_1}A^{i_2}\cdots A^{i_N})|i_1,\ldots,i_N\rangle,
\end{equation}
where $i_j$ labels 
the local basis states at site $j$, $A$ is a rank-3 tensor with $A^{i_j}$ a matrix in the virtual space, and "$\text{tr}$" ("$\text{Tr}$") denotes tracing over the virtual (physical) indices. We choose $X=\mathbb{1}$ (periodic boundary condition) to represent the state $|\psi\rangle\equiv|\psi(\mathbb{1})\rangle$; other boundary conditions will be used later to probe the SPT order.

The characteristic feature of SPT states in the MPS representation is the symmetry transformation rule of the tensor $A$. In particular, $A$ can be chosen to satisfy~\cite{garcia2008prl,cirac2021rmp} 
\begin{equation}
\sum_j u_{g}^{ij}A^{j}=e^{i\theta_g}V_gA^{i}V_g^{\dagger}.
\label{eq:injectivity_pure}
\end{equation}
Here, $V_g$ forms a projective representation of the symmetry group, i.e., $V_{g_1}V_{g_2}=V_{g_1g_2}\omega(g_1,g_2)$, with  $\omega$ being a phase factor \cite{chen2013prb}. The choice of the $V_g$ and $\omega$ is not unique, due to the gauge transformations: $V_g\rightarrow V_ge^{i\phi(g)}$ and $\omega(g_1,g_2)\rightarrow \omega(g_1,g_2)e^{i\phi(g_1 g_2)-i\phi(g_1)-i\phi(g_2)}$. This redundancy defines equivalence classes, classified by the group cohomology~\cite{chen2011prb, schuch2011prb, chen2013prb}. 

The transformation rule~\eqref{eq:injectivity_pure} ensures symmetry under $U_g$ in the bulk for $|\psi(X)\rangle$,
\begin{equation}\label{eq:charge_extensive}
U_g|\psi(X)\rangle=e^{iN\theta_g}|\psi(V_g^{\dagger}XV_{g})\rangle,
\end{equation}
while the boundary lacks this symmetry unless $V_g^\dagger X V_g \propto X$ (up to a phase factor), such as when $X=\mathbb{1}$.
 This property can be utilized to detect SPT phases through the $g_2$-symmetry charge induced from a $g_1$-twisted boundary condition, where $g_1$ and $g_2$ denote two commuting group elements hereafter.  Specifically, taking $X=V_{g_1}$, the boundary is also $g_2$-symmetric since $V_{g_2}^\dagger V_{g_1}V_{g_2} =[\omega(g_1, g_2)/\omega(g_2, g_1)]V_{g_1}$, yielding
\begin{equation}
U_{g_2}|\psi(V_{g_1})\rangle=e^{iN\theta_{g_2}}\frac{\omega(g_1,g_2)}{\omega(g_2,g_1)}|\psi(V_{g_1})\rangle.
\label{eq:symm_transform}
\end{equation}
Physically,
$|\psi(V_{g_1})\rangle$ is the $g_1$-flux inserted MPS state~\cite{zaletel2013prl,zaletel2014njp}. The extensive phase factor $e^{iN\theta_{g_2}}$ represents the symmetry charge without flux insertion [Eq.~\eqref{eq:charge_extensive}], while the ratio $\omega(g_1,g_2)/\omega(g_2,g_1)$ captures the additional quantized topological $g_2$-charge induced by the $g_1$-flux insertion, i.e.,
\begin{equation}
\frac{\omega(g_1,g_2)}{\omega(g_2,g_1)}=\frac{\langle \psi(V_{g_1})|U_{g_2}|\psi(V_{g_1})\rangle}{\langle \psi|U_{g_2}|\psi\rangle}.
\label{eq:pure_quantized}
\end{equation}
The response is quantized because the phase accumulates multiplicatively with each additional $g_1$ flux, and adding $|g_1|$ (the order of $g_1$, i.e., $g_1^{|g_1|}=\mathbb{1}$) fluxes results in a trivial response \cite{zaletel2014njp}.

\textcolor{red}{\emph{Quantized response in mixed states.--}} We now show that the above discussion extends to mixed states. We will focus on those described by LPDOs, with purifications admitting short-range correlated MPS representations~\cite{verstrate2004prl, zwolak2004prl, cuevas2013njp},
\begin{equation}
|\Psi(X)\rangle \!=\!\sum_{\{i_j ,a_j\}}\text{tr}(X A^{i_1 a_1}\cdots  A^{i_N a_N})|i_1a_1,\ldots,i_Na_N\rangle.
\end{equation}
Here, $i_j$ and the additional index $a_j$ label the physical and ancilla states at site $j$, respectively. Similar to the pure-state case, we introduce a boundary matrix $X$ acting on the virtual space. For later convenience, we define the following graphical notation for $(A^{ia})_{\alpha \beta}$: $\begin{array}{c} 
    \begin{tikzpicture}[scale=.4, baseline={([yshift=-5ex]current bounding box.center)}, thick]
    	\ATensorLPDO{0,0}{$A$}
    	\draw (-0.4,1.2) node {\scriptsize $i$};
	\draw (0.9,1.2) node {\scriptsize $a$};
    	\draw (1.8,0) node {\scriptsize $\beta$};
    	\draw (-1.8,0) node {\scriptsize $\alpha$};
    \end{tikzpicture}
\end{array},$ where the red leg represents ancillas, and connecting legs of tensors represents the contraction of corresponding indices in later formulas. The system density matrix is obtained by tracing over the ancillas, i.e., $\rho(X)=\text{Tr}_a|\Psi(X)\rangle\langle\Psi(X)|$, with $\rho(\mathbb{1})\equiv\rho$.

Symmetries in mixed states, unlike the pure state case, bifurcate into two classes \cite{buca2012njp}: a symmetry $g$ is weak for $\rho$ if $\left[U_{g}, \rho\right]=0$; and strong if $U_{g}\rho\sim \rho$ up to a phase factor. 
We consider symmetries which admit a local realization in the extended Hilbert space including ancillas, i.e., $U_g\otimes U_g^a= (u_g\otimes u^a_g)^{\otimes N}$~\cite{degroot2022quantum}, which represents the symmetry of the purified MPS of $\rho$. Similarly to Eq.~\eqref{eq:injectivity_pure},  with a proper gauge choice, $u_g\otimes u_g^a$ acts on $A^{ia}$ as
\begin{equation}\label{eq:injectivity_weak}
\begin{array}{c}
\begin{tikzpicture}[scale=.5,thick,baseline={([yshift=-9ex]current bounding box.center)}]
			\ATensorLPDO{0,0}{$A$}
			\uTensor{-0.6,0.9}{$u_g$}
            \uTensorA{0.6,0.9}{$u_g^a$}
\end{tikzpicture}
\end{array}
=e^{i\theta_g}
\begin{array}{c}
\begin{tikzpicture}[scale=.5,thick,baseline={([yshift=-4ex]current bounding box.center)}]
			\ATensorLPDO{0,0}{$A$}
			\vTensor{-2.2,0}{$V_g$}
            \vTensor{1.2,0}{$V_g^\dagger$}
\end{tikzpicture}
\end{array}.
\end{equation}
 The key distinction between these two symmetries lies in $u_g^a$: The strong one requires the ancilla to be  $g$-charge neutral ($u_g^a=\mathbb{1}$), whereas the weak one imposes no such restriction. We focus on the weak symmetry case hereafter, where the weak $g_1$-flux inserted state~\cite{suppupdated}, obtained by taking $X=V_{g_1}$ and denoted as $\rho(V_{g_1})$, preserves the $g_2$ weak symmetry. 

Now we define the $g_2$-charge response from the $g_1$-flux insertion, and demonstrate its quantization, as it remains a multiplicative phase factor upon adding $g_1$-fluxes. Here unlike the pure state (or strong symmetry) case, the $g_2$ symmetry holds only on average: Each eigenstate of the density matrix is a $g_2$-symmetric but can have distinct symmetry charges. Correspondingly, in the $N\gg1$ limit, the ensemble average of $U_{g_2}$ decays with system size due to interference,
\begin{align}
\text{Tr}[\rho(V_{g_1})U_{g_2}]\sim e^{-\Theta_{g_2} N+i\mathcal{Q}(g_1, g_2)}.
\end{align} 
As demonstrated below, the extensively decaying amplitude arises from non-universal bulk contributions, independent of the boundary condition $X=V_{g_1}$, whereas the phase term $e^{i \mathcal{Q}(g_1,g_2)}$ captures topological information. These results follow from the tensor diagrams below,  
\begin{equation}
\begin{split}
\text{Tr}[\rho(V_{g_1})U_{g_2}]&=\begin{array}{c}
	\begin{tikzpicture}[scale=.5, baseline={([yshift=9ex]current bounding box.center)}, thick]
		\draw[shift={(0,0)},dotted] (0,0) -- (6,0);
        \draw[shift={(0,0)},dotted] (0,2.8) -- (6,2.8);
        \draw[shift={(0,0)}] (-2.2,0) arc(90:270:0.2);
        \draw[shift={(0,2.8)}] (-2.2,0) arc(90:270:0.2);
        \draw[shift={(0,0)}] (6.6,-0.4) arc(-90:90:0.2);
        \draw[shift={(0,0)}] (6.6,2.4) arc(-90:90:0.2);
		\ATensorLPDO{0.4,0}{$A$}
        \ADaggTensorLPDO{0.4,2.8}{$A^*$}
        \uTensor{-0.2,0.9}
        {$u_{g_2}$}
        \vTensor{-1.9,0}{$V_{g_1}$}
		\ATensorLPDO{5,0}{$A$}
        \ADaggTensorLPDO{5,2.8}{$A^*$}
        \uTensor{4.4,0.9}{$u_{g_2}$}
        \vTensor{-1.9,2.8}{$V_{g_1}^*$}
		\draw [decorate,
    	decoration = {calligraphic brace,mirror}] (-0.5,-0.8) --  (5.9,-0.8);
		\draw (2.7,-1.5) node {\scriptsize $N$};
        \draw[red](1,0.7)--(1,2);
        \draw[red](5.6,0.7)--(5.6,2);
	\end{tikzpicture}
	\end{array}\\
    &=\text{tr}[(V_{g_1}^{*}\otimes V_{g_1})T(g_2)^N],
    \end{split}
    \label{eq:symm_charge}
\end{equation}
which is contracted horizontally. The $g_2$-twisted transfer matrix $T(g_2)$ can be written as
\begin{equation}
\begin{split}
T(g_2)&\equiv
\begin{array}{c}
	\begin{tikzpicture}[scale=.5, baseline={([yshift=0ex]current bounding box.center)}, thick]
		\ATensorLPDO{0.4,0}{$A$}
        		\ADaggTensorLPDO{0.4,2.8}{$A^*$}
       		 \uTensor{-0.2,0.9} {$u_{g_2}$}
		  \draw[red](1,0.7)--(1,2);
	\end{tikzpicture}
\end{array}= \lambda^{(0)}_{g_2}
	\begin{array}{c}
		\begin{tikzpicture}[scale=.5, baseline={([yshift=0ex]current bounding box.center)}, thick]
		\draw[rounded corners] (0,2.8)--(0.5,2.8)--(0.5,0)--(0,0);
		\uTensor{0.5,0.9}{$R_{g_2}^0$}
		\draw[rounded corners] (2.2,2.8)--(1.7,2.8)--(1.7,0)--(2.2,0);
		\uTensor{1.7,0.9}{$L_{g_2}^0$}
		\end{tikzpicture}
	\end{array}+\delta T(g_2)\\
&=\lambda_{g_2}^{(0)}|R_{g_2}^0)(L_{g_2}^0|+\delta T(g_2).
\quad
\end{split}
\label{eq:transfer_matrix}
\end{equation}
Here, $\lambda_{g_2}^{(0)}$ is the largest eigenvalue of $T(g_2)$ in magnitude, assumed to be non-degenerate. The corresponding right and left eigenvectors $|R_{g_2}^0)$ and $(L_{g_2}^0|$ are normalized (i.e., $(L_{g_2}^0|R_{g_2}^0)=1$). The remainder $\delta T(g_2)$ has spectrum radius $|\lambda_{g_2}^{(1)}|<|\lambda_{g_2}^{(0)}|$, with $\lambda_{g_2}^{(1)}$ the subleading eigenvalue in magnitude, indicating a finite spectral gap in $T(g_2)$,
\begin{equation}
\Delta(g_2)\equiv |\lambda_{g_2}^{(0)}|-|\lambda_{g_2}^{(1)}|>0,
\label{eq:gap}
\end{equation}
referred to as the $g_2$-symmetry gap hereafter.

The $g_2$-symmetry gap protects the quantized response, given by
\begin{equation}
e^{i\mathcal{Q}(g_1,g_2)}=\frac{\text{Tr}[\rho(V_{g_1})U_{g_2}]}{\text{Tr}(\rho U_{g_2})}
=\frac{\text{tr}[(V_{g_1}^*\otimes V_{g_1})T(g_2)^N]}{\text{tr}[T(g_2)^N]}.
\end{equation}
The denominator $\text{Tr}(\rho U_{g_2})$ removes the extensive factor $e^{-\Theta_{g_2}N}=[\lambda_{g_2}^{(0)}]^N$ that is independent of $g_1$. Crucially, $\Delta(g_2)$ ensures that, in the $N\gg 1$ limit, $\frac{T(g_2)^N}{\text{tr}[T(g_2)^N]}$ approaches $|R_{g_2}^0)(L_{g_2}^0|$, 
a projector onto the leading eigenvector of $T(g_2)$. This yields
\begin{equation}
e^{i\mathcal{Q}(g_1,g_2)}=(L_{g_2}^{0}|(V_{g_1}^*\otimes V_{g_1})|R_{g_2}^0),
\end{equation}
expressing $e^{i\mathcal{Q}(g_1,g_2)}$ as a single quantum amplitude, allowing the standard quantization argument used in pure-state cases: $(V_{g_1}^{*}\otimes V_{g_1})$ is a symmetry of $T(g_2)$ (see Appendix \hyperref[sec:commutation]{A} for a diagrammatic proof),
so $e^{i\mathcal{Q}(g_1,g_2)}$ is an eigenvalue of $V_{g_1}^*\otimes V_{g_1}$, associated with the eigenvector $|R_{g_2}^0)$. We thus confirm that the $g_2$-charge response is multiplicative upon inserting multiple $g_1$-fluxes, and, moreover, remains quantized for any finite group $G$.

Changing the quantized response $e^{i\mathcal{Q}(g_1,g_2)}$ requires closing the $g_2$-symmetry gap, whereas notably, long-range correlations in local operators are generally not required -- a point we will explore later through a concrete model. Furthermore,  unlike pure states where $e^{i\mathcal{Q}(g_1,g_2)}$ is given by the group cohomology invariant $\omega(g_1,g_2)/\omega(g_2,g_1)$ [see Eq.~\eqref{eq:pure_quantized}], we will show that weakly symmetric open systems can exhibit different values, suggesting the need for a more general mathematical framework. Finally, while the LPDO discussion does not require strong injectivity, we note that our discussion naturally applies to states described by strong injective matrix product density operators~\cite{guo2024arxiv_lpdo, xue2024arxiv_mpdo}  (MPDOs) as well. For these states, weak symmetries can be pushed to virtual space similarly to Eq.~\eqref{eq:injectivity_weak},  and the above discussion remains applicable replacing $V_{g_1}^{*}\otimes V_{g_1}$ with the corresponding transformations in the virtual space.

\textcolor{red}{\emph{String order parameters.--}} One might be concerned that the quantized phase above is a ratio of two exponentially small numbers in system size, making it challenging to measure. To address this practical issue, we propose using string order parameters \cite{nijs1989prb,kennedy1992prb,garcia2008prl,pollmann2012prb} on finite segments as probes of $\mathcal{Q}(g_1,g_2)$ in a large system: 
\begin{equation}\label{eq:string_original}
\mathcal{S}(g_2,\chi^L_{g_1},\chi^R_{g_1})=\text{Tr}\left[\rho\ \chi^L_{g_1}\otimes\left(\otimes_{i=j}^{j+l-1} u_{g_2} \right)\otimes\chi^R_{g_1}\right].
\end{equation}
Here, $\left(\otimes_{i=j}^{j+l-1} u_{g_2}\right)$ creates a symmetry-twisted domain of length $l$. The operators $\chi^{L/R}_{g_1}$, separating twisted and untwisted domains, carry a $g_1$-charge, 
\begin{equation}\label{eq:chi_LR}
u_{g_1}\chi_{g_1}^{L/R} u_{g_1}^\dagger = e^{\pm i \phi(g_1)}\chi_{g_1}^{L/R}.
\end{equation}  
In the pure-state limit (i.e., for ordinary SPT states), $\mathcal{Q}(g_1,g_2)$ is associated with the group cohomology invariant $\omega(g_1,g_2)/\omega(g_2,g_1)$, known to be captured by the dependence of the string order parameter on $\phi(g_1)$, as reflected in the corresponding selection rule~\cite{pollmann2012prb}. 

For general cases, the topological charge $\mathcal{Q}(g_1,g_2)$ reveals itself analogously. Tuning the parameter $\phi(g_1)$ by varying $\chi_{g_1}^{L/R}$ results in distinct decay behaviors of the string order parameter with string length $l$: Scaling behavior $\mathcal{S}(g_2,\chi_{g_1}^L,\chi_{g_1}^R)\sim [\lambda_{g_2}^{(0)}]^l$ occurs only if $e^{i\mathcal{Q}(g_1,g_2)}=e^{i\phi(g_1)}$, while other choices yield subleading scaling, e.g., $[\lambda_{g_2}^{(1)}]^l$, as demonstrated below with a concrete example, and proven in \cite{suppupdated}.
To illustrate this behavior, we introduce a normalized string order parameter by dividing out the extensive part that leads to exponential decay,
\begin{equation}
\mathcal{S}^{(n)}(g_2,\chi_{g_1}^L,\chi_{g_1}^R)=\frac{\mathcal{S}(g_2,\chi_{g_1}^L,\chi_{g_1}^R)}{|\text{Tr}\left(\rho U_{g_2}\right)|^{l/N}},
\label{eq:normalized_string}
\end{equation}
satisfying the following selection rule,
\begin{equation}\label{eq:selection_normalized_sop}
\mathcal{S}^{(n)}(g_{2},\chi_{g_{1}}^{L},\chi_{g_{1}}^{R})|_{l\rightarrow\infty}=\begin{cases}
\mathcal{O}\left(1\right),\ \text{if}\ e^{i\mathcal{Q}(g_{1},g_{2})}= e^{i\phi(g_{1})}\\
0,\quad\text{if}\ e^{i\mathcal{Q}(g_{1},g_{2})}\neq e^{i\phi(g_{1})}
\end{cases}.
\end{equation}
Besides open quantum systems, similar ideas dealing with exponentially decaying string order parameter,
associated with emergent higher-form symmetries, have been used to define the Fredenhagen-Marcu string order parameters~\cite{fredenhagen1983cmp,marcu1986springer,xu2024arxiv} for detecting topological order. 
    Additionally, this construction parallels earlier probes of bulk topology and boundary anomaly using defect operators ~\cite{huang2022prb,jiang2023scipost,huang2024arxiv_mstop, huang2025prl,wang2024prl, zang2024prl, mao2024arxiv}, and aligns with results derived from field theory in open quantum systems~\cite{huang2025prl}.

\textcolor{red}{\emph{Example: AKLT state.--}}
The key concept that allows us to define phases and phase transitions is the $g_2$-symmetry gap [Eq.~\eqref{eq:gap}], i.e., the spectral gap of the twisted transfer matrix $T(g_2)$. Meanwhile, the correlation length for local observables is set by the gap of $T(\mathbb{1})$, whose closing signals long-range correlations. In the strong symmetry case, $T(g_2)$ and $e^{i\theta_{g_2}}T(\mathbb{1})$ are related by similarity transformations which preserve the spectrum. Consequently, all $g_2$-symmetry gaps open and close at the same time, and thus the phase transition point must develop long-range correlations in the state. By contrast, for weak symmetries, this condition no longer holds. A phase transition can occur by closing one of the $g_2$-symmetry gaps while $T(\mathbb{1})$ remains gapped, indicating finite correlation length for local observables as anticipated above. In the following, we will demonstrate this physics by constructing a prototype example. 

To generate an LPDO with a short-range correlated purification, as considered within our framework, one representative way is to apply local noise channels to short-range correlated MPSs. 
We consider the pure spin-1 AKLT state described by the following MPS tensor~\cite{pollmann2017oxford, cirac2021rmp}, i.e.,
\begin{equation}
A^{-1}=\sqrt{\frac{2}{3}}\sigma^{+},\ \  A^0=-\sqrt{\frac{1}{3}}\sigma^z,\ \   A^{+1}=-\sqrt{\frac{2}{3}}\sigma^{-},
\label{eq:AKLT_tensor}
\end{equation}
where we use standard notation for the Pauli matrices, and the physical indices  label local spin-z basis states.
This is the ground state of the AKLT model, and has mutually commuting $\mathbb{Z}_2$ symmetries generated by global $\pi$-rotations of spin along different axes, 
represented as $R_\alpha^{\otimes N}$. Here, $R_\alpha \equiv  e^{i\pi S_\alpha}$, and $S_\alpha \in \{S_0,\ S_x,\ S_y,\ S_z\}$ with $S_0$ being the identity matrix and $S_{x,y,z}$ the usual spin-1 operators. 
Applying onsite noisy channels to the AKLT state $\rho_{0}$, we obtain
\begin{equation}
\rho=\mathcal{N}_1\circ\mathcal{N}_2\circ\cdots\mathcal{N}_N[\rho_{0}],\ \text{and}\  \mathcal{N}_i[\cdot]=\sum_{\alpha} K_{\alpha,i}(\cdot) K_{\alpha,i}^{\dagger},
\label{eq:on-site-channel}
\end{equation}
where the Kraus operators $\{K_{\alpha,i}\}$ of $\mathcal{N}_i$ act locally on the spin-$1$ Hilbert space at site-$i$ as $\{\sqrt{1-p}S_0,\sqrt{p}S_x S_y, \sqrt{p}S_yS_z,\sqrt{p} S_z S_x\}$, with $p\in [0,1]$ representing the noise rate. The noise breaks the AKLT state's strong $\mathbb{Z}_2$ symmetries down to weak ones. The resulting state $\rho$ can be locally purified to a short-range correlated state by considering Stinespring's dilation~\cite{nielsen2010cambridge} for each onsite channel. 

\begin{figure}[t]  
\centering
\includegraphics[width=1\linewidth]{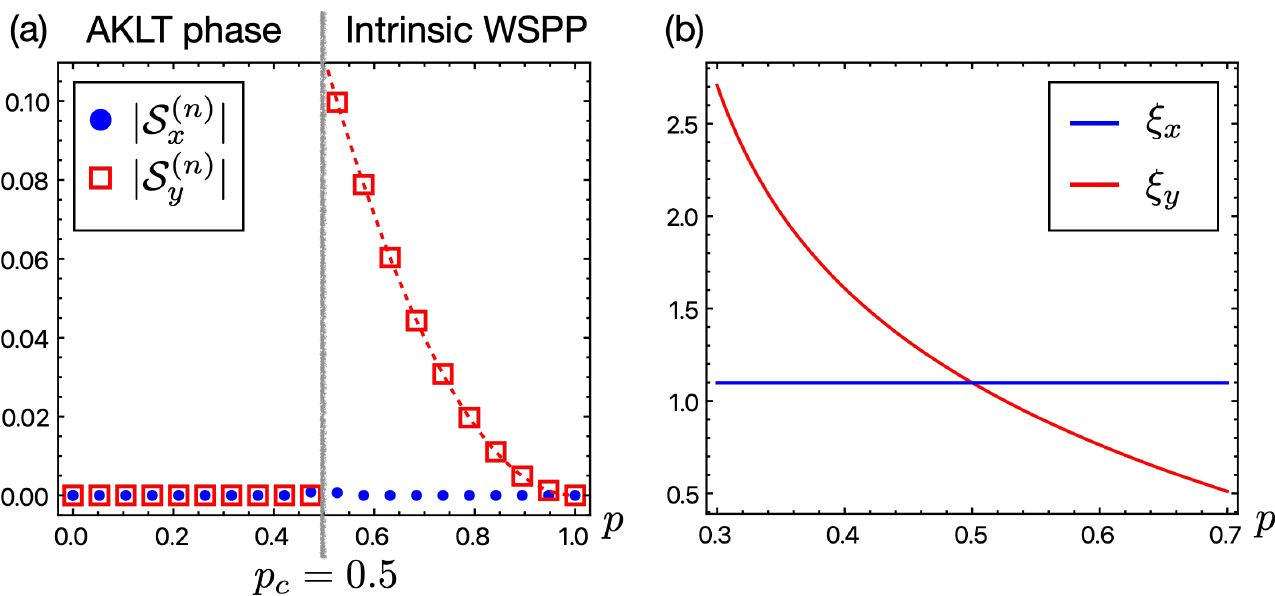}

\caption{Phase detection for the decohered AKLT state at different noise rate $p$  via string order parameters. Panel (a) shows the normalized string order parameters in a 200-site periodic chain with string length 50. The dashed line is analytically computed in the limit of $N\gg l\gg 1$. The quantized responses $(e^{i\mathcal{Q}(R_x, R_z)},e^{i\mathcal{Q}(R_y, R_z)})$ identify two phases: $(-1,\ -1)$, representing the AKLT phase, and $(-1,\ +1)$, an intrinsic weak-symmetry protected phase (WSPP). These phases manifest as distinct patterns in the normalized string order parameter, with $\mathcal{S}_y^{(n)}=0$ for the former, and $\mathcal{O}(1)$ for the latter.  Panel (b) presents the analytically computed decay exponents $\xi_x$ and $\xi_y$ (cf. Eq.~\eqref{eq: string}); their crossing point marks the transition~\cite{fn_fig2}.\label{fig:AKLT_string} }
\end{figure}

 Below we summarize key findings, while a detailed analysis is provided in Appendix \hyperref[sec:AKLT]{B}. This model exhibits two distinct phases that can be characterized by quantized responses $e^{i\mathcal{Q}(R_x, R_z)}$ and $e^{i\mathcal{Q}(R_y, R_z)}$.  The conventional AKLT phase appears in the low noise regime ($p<1/2$), with $\left(e^{i\mathcal{Q}(R_x, R_z)}, e^{i\mathcal{Q}(R_y, R_z)}\right)=(-1,\ -1)$. As the noise rate increases ($p>1/2$), a new phase emerges, characterized by $(-1,\ +1)$. This phase intrinsically relates to weak symmetry: The sign change in $e^{i\mathcal{Q}(R_y, R_z)}$ signals a switch of the leading eigenvector in $T(R_z)$, accompanied by the closing of the $R_z$-symmetry gap while preserving the $T(\mathbb{1})$ gap - a phenomenon impossible in pure-state or strong symmetry cases, which in turn gives rise to a quantized response beyond the group cohomology description. Moreover, these quantized charges result in qualitatively distinct behavior of the string order parameters [cf. Eq.~\eqref{eq:string_original}], expressed as
\begin{equation}
\mathcal{S}_\alpha \equiv \mathcal{S}(R_z, S_{\alpha},S_{\alpha}) =\text{Tr}\left[\rho S_\alpha \otimes \left( \otimes_{i=j}^{j+l-1}R_z\right)\otimes S_\alpha\right],
\label{eq:string_AKLT}
\end{equation}
and their normalized counterparts $\mathcal{S}_\alpha^{(n)}$.
Figure \ref{fig:AKLT_string}(a) shows that $\mathcal{S}^{(n)}_y$ distinguishes two phases: It vanishes in the $(-1,\ -1)$ phase due to the selection rule [Eq.~\eqref{eq:selection_normalized_sop}], but reaches an order-one value in the  weak-symmetry protected phase. Conversely, $\mathcal{S}^{(n)}_x$ vanishes in both phases.

In practice, we can probe this transition by contrasting qualitatively different decay behaviors of string order parameters $\mathcal{S}_\alpha$ as functions of string length. 
These behaviors are characterized by the relevant decay exponents, defined as
\begin{equation} 
\begin{split}
\xi_{\alpha} &=-\lim_{l\rightarrow \infty}\frac{1}{l}\log|\mathcal{S}_\alpha|.
\end{split}
\label{eq: string}
\end{equation}
Here, the $\alpha$ dependence of $\xi_\alpha$ reflects the selection rule, as $\xi_\alpha$ asymptotically converges to the $T(R_z)$ eigenvalue with dominant contribution to $\mathcal{S}_\alpha$. This is confirmed by the analytical results plotted in Fig.~\ref{fig:AKLT_string}(b).

\textcolor{red}{\emph{Discussion and outlook.--}} Our work demonstrates that weak symmetries can protect topological responses in mixed-state bosonic systems, extending beyond the strong-symmetry requirement discussed in previous studies. 
Our approach differs from previous ones extending the circuit approach for pure states to define mixed-state phases via local quantum channels~\cite{ruiz2024lmp} or fast local Lindbladian evolution~\cite{coser2019quantum}, where existing nontrivial generalizations~\cite{ma2023prx, degroot2022quantum} of bosonic SPT phases rely on strong symmetry. This distinction arises from different perspectives in generalizing from the pure-state cases, where SPT phases can be described by their topological responses as order parameters~\cite{hassan2010rmp,qi2011rmp,pollmann2012prb,zaletel2014njp}, or more formally with the circuit approach, as equivalence classes under symmetric local unitary transformations~\cite{chen2013prb}. In mixed states, these approaches lead to different notions of phases. The order-parameter approach reveals a weak-symmetry protected phase, characterized by quantized responses and the selection rule for string order parameters (see Fig.~\ref{fig:AKLT_string}). This notion of phase has a clear operational meaning and is detectable in current experimental platforms \cite{endres2011science, semeghini2021science,iqbal2024nature, karch2025arxiv}.

Central to our findings is the symmetry gap of the symmetry-twisted transfer matrix, which ensures the robustness of topological responses. Consequently, phase transitions occur via symmetry gap closures without standard thermodynamic (local) signatures, a situation well-described within the tensor network framework and amenable to efficient numerical simulations. Similar patterns are observed in fermionic systems and the decohered toric code~\cite{fan2024prxQ, huang2025prl, huang2024CI}, where topological phase transitions have been described as the loss of topological modes~\cite{huang2025prl, huang2024CI}.
Future directions include extending the tensor-network approach to higher-dimensional and fermionic systems, and developing a systematic framework for classifying mixed-state phases protected by weak symmetry and its symmetry gap. These may ultimately yield a unified understanding of mixed-state topological phases, with implications for quantum error correction~\cite{fan2024prxQ, lee2023prxq, huang2024CI, behrends2024arxiv}.

\begin{acknowledgments}
We thank Ignacio Cirac for helpful discussions.
Z.-M.~H. and S.~D. are supported by the  Deutsche Forschungsgemeinschaft (DFG, German Research Foundation) under Germany’s Excellence Strategy Cluster of Excellence Matter and Light for Quantum Computing (ML4Q) EXC 2004/1 390534769 and by the DFG Collaborative Research Center (CRC) 183 Project No. 277101999 - project B02. X.-Q.~S. acknowledges  support from the Deutsche Forschungsgemeinschaft (DFG, German Research Foundation) under
Germany’s Excellence Strategy -- EXC-2111 -- 390814868 and from the Alexander von Humboldt Foundation.
\end{acknowledgments}

\appendix
\clearpage
\section*{End Matter}
\section{Appendix A: Proof of $[V_{g_1}^*\otimes V_{g_1},T(g_2)]=0$}
\label{sec:commutation}
We present a diagrammatic proof of the identity,
\begin{equation}\label{supp_eq:sym_flux}
    [V_{g_1}^*\otimes V_{g_1},T(g_2)]=0, \text{\ for\ commuting $g_1,g_2$, }
\end{equation}
or equivalently,
\begin{equation}
T(g_2)=\left(V_{g_1}^*\otimes V_{g_1}\right)\ T(g_2)\ \left(V_{g_1}^T\otimes V^{\dagger}_{g_1}\right).
\end{equation}
The proof directly follows from the tensor network diagram below,
\begin{equation}
\begin{split}
\begin{array}{c}
	\begin{tikzpicture}[scale=.5, baseline={([yshift=0ex]current bounding box.center)}, thick]
		\ATensorLPDO{0.4,0}{$A$}
        		\ADaggTensorLPDO{0.4,2.8}{$A^*$}
       		 \uTensor{-0.2,0.9} {$u_{g_2}$}
		  \draw[red](1,0.7)--(1,2);
	\end{tikzpicture}=
    	\begin{tikzpicture}[scale=.5, baseline={([yshift=0ex]current bounding box.center)}, thick]
		\ATensorLPDO{0.4,-1.3}{$A$}
        \ADaggTensorLPDO{0.4,4.1}{$A^*$}
          \draw[red](1,0.7)--(1,2);
             \uTensor{-0.2,-0.4} {$u_{g_1}$}
       		 \uTensor{-0.2,0.9} {$u_{g_2}$}
             \uTensor{-0.2,2.2} {$u_{g_1}^{\dagger}$}
             \uTensorA{1,-0.4} {$u_{g_1}^{a}$}
             \uTensorA{1,2.2} {$u_{g_1}^{a\dagger}$}
	\end{tikzpicture}
    =\begin{tikzpicture}[scale=.5, baseline={([yshift=0ex]current bounding box.center)}, thick]
            \ATensorLPDO{0.4,0}{$A$}
            \ADaggTensorLPDO{0.4,2.8}{$A^*$}
            \vTensor{-1.8,0}{$V_{g_1}$}
            \vTensor{1.6,0}{$V_{g_1}^{\dagger}$}
            \vTensor{-1.8,2.8}{$V_{g_1}^{*}$}
            \vTensor{1.6,2.8}{$V_{g_1}^{T}$}
       		\uTensor{-0.2,0.9} {$u_{g_2}$}
		  \draw[red](1,0.7)--(1,2);
	\end{tikzpicture}
\end{array}
\end{split}.
\end{equation}
The first equality rewrites $u_{g_2}$ as  $u_{g_1}^{\dagger}u_{g_2}u_{g_1}$ for commuting $g_1,\ g_2$, and then introduces $u_{g_1}^a$ on the ancilla line using the identity $u_{g_1}^{a\dagger}u_{g_1}^a=\mathbb{1}$. The second equality follows from applications of the weak symmetry transformation rule of $A$ [see Eq.~\eqref{eq:injectivity_weak}].

\appendix 
\section{Appendix B: Analytical results of the decohered AKLT state}
\label{sec:AKLT}
We present analytical results for the AKLT state under decoherence, including the spectrum of the (symmetry-twisted) transfer matrix, the quantized response, and the string order parameters.



\begin{table}[htbp]

\begin{tabular}{|c|c|c|c|}
\hline 
Eigenvalue & Eigenvector  & $e^{iq_x}$ & $e^{iq_y}$
\tabularnewline
\hline 
\hline 
$1$ & $\frac{1}{\sqrt{2}}\left(1,\ 0,\ 0,\ 1\right)^{T}$ & $1$ & $1$\tabularnewline
\hline 
$-\frac{1}{3}$ & $\frac{1}{\sqrt{2}}\left(-1,\ 0,\ 0,\ 1\right)^{T}$ & $-1$ & $-1$ \tabularnewline
\hline 
$-\frac{1}{3}$ & $\frac{1}{\sqrt{2}}\left(0,\ 1,\ 1,\ 0\right)^{T}$ & $1$ & $-1$\tabularnewline
\hline 
$-\frac{1}{3}$ & $\frac{1}{\sqrt{2}}\left(0,\ -1,\ 1,\ 0\right)^{T}$ & $-1$ & $1$\tabularnewline
\hline 
\end{tabular}%
\caption{Summary of the eigenvalues and
eigenvectors of $T(\mathbb{1})$,  together with the corresponding $V_{R_{\mu}}^*\otimes V_{R_{\mu}}$ eigenvalues $e^{iq_\mu}$ ($\mu=x, y$). \label{supp_table_T_notwist}}
\end{table}

\begin{table}[htbp]
\begin{tabular}{|c|c|c|c|}
\hline 
Eigenvalue & Eigenvector  & $ e^{iq_x}$ & $e^{iq_y}$\tabularnewline
\hline 
\hline 
$\frac{3-4p}{3}$ & $\frac{1}{\sqrt{2}}\left(-1,\ 0,\ 0,\ 1\right)^{T}$ & $-1$ & $-1$\tabularnewline
\hline 
$\frac{-1+4p}{3}$ & $\frac{1}{\sqrt{2}}\left(0,\ -1,\ 1,\ 0\right)^{T}$ & $-1$ & $1$\tabularnewline
\hline 
$-\frac{1}{3}$ & $\frac{1}{\sqrt{2}}\left(1,\ 0,\ 0,\ 1\right)^{T}$ & $1$ & $1$\tabularnewline
\hline 
$-\frac{1}{3}$ & $\frac{1}{\sqrt{2}}\left(0,\ 1,\ 1,\ 0\right)^{T}$ & $1$ & $-1$\tabularnewline
\hline 
\end{tabular}\caption{Summary of the eigenvalues and
eigenvectors of $T(R_z)$,  together with the corresponding $V_{R_{\mu}}^*\otimes V_{R_{\mu}}$ eigenvalues $e^{iq_\mu}$ ($\mu=x, y$). 
\label{supp_table_T_twist}}
\end{table}


Building on the MPS representation of the AKLT state [Eq.~\eqref{eq:AKLT_tensor}], we first calculate spectra of the transfer matrices with and without a symmetry twist, namely, $T(\mathbb{1})$ and $T(R_z)$, for the decohered state obtained after applying on-site noisy channels [Eq.~\eqref{eq:on-site-channel}] for different noise rate $p$. 
Specifically for $T(\mathbb{1})$, the spectrum is unchanged under decoherence due to the trace-preserving nature of the on-site channels:
\begin{equation}
T(\mathbb{1}) = \frac{1}{3}\left(\begin{array}{cccc}
1 & 0 & 0 & 2\\
0 & -1 & 0 & 0\\
0 & 0 & -1 & 0\\
2 & 0 & 0 & 1
\end{array}\right),
\end{equation}
whose eigenvalues and eigenvectors are shown in Table \ref{supp_table_T_notwist}. By contrast, the spectrum of $T(R_z)$ depends on $p$:
\begin{eqnarray}
&&T(R_z)\nonumber\\
&=&\left(\begin{array}{cccc}
\frac{1}{3}\left(1-2p\right) & 0 & 0 & -\frac{2}{3}\left(1-p\right)\\
0 & \frac{1}{3}\left(-1+2p\right) & -\frac{2p}{3} & 0\\
0 & -\frac{2}{3}p & \frac{1}{3}\left(-1+2p\right) & 0\\
-\frac{2}{3}\left(1-p\right) & 0 & 0 & \frac{1}{3}\left(1-2p\right)
\end{array}\right),\nonumber\\
\end{eqnarray}
with eigenvalues and eigenvectors also shown in Table \ref{supp_table_T_twist}. 

In the thermodynamic limit,  the quantized response [namely, $e^{i\mathcal{Q}(R_{\mu},R_z)}$ with $\mu=x, y$] is determined by the leading eigenvector of $T(R_z)$, with its value given by the corresponding $V_{R_{\mu}}^* \otimes V_{R_{\mu}}$ eigenvalue. This can be directly read from Table~\ref{supp_table_T_twist}: 
\begin{itemize}
    \item For $p<\frac{1}{2}$, the leading eigenvector is $\frac{1}{\sqrt{2}}(-1,\ 0,\ 0,\ 1)^T$, with a quantized response $(e^{i\mathcal{Q}(R_x,\ R_z)},\ e^{i\mathcal{Q}(R_y,\ R_z)})=(-1,\ -1)$, belonging to the same phase as the AKLT state.
    \item For $p>\frac{1}{2}$, the leading eigenvector becomes $\frac{1}{\sqrt{2}}(0,\ -1,\ 1,\ 0)^T$, and the quantized response shifts to $(e^{i\mathcal{Q}(R_x,\ R_z)},\ e^{i\mathcal{Q}(R_y,\ R_z)})=(-1,\ +1)$.
\end{itemize}
The transition occurs through an exchange of leading eigenvectors, with the critical point at $p=\frac{1}{2}$, where the $R_z$-symmetry gap closes.

We now consider the string order parameters $\mathcal{S}_{\alpha}$ [Eq.~\eqref{eq:string_AKLT}] and their normalized counterparts $\mathcal{S}_{\alpha}^{(n)}$ in the limit of $N\gg l\gg 1$, which can be computed knowning the eigenvectors and eigenvalues of $T(\mathbb{1})$ and $T(R_z)$. 
In this limit, the leading contributions to the normalized string order parameters $\mathcal{S}_x^{(n)}$ and $\mathcal{S}_y^{(n)}$, as discussed in the main text (cf. Fig.~\ref{fig:AKLT_string}),  are given by:
\begin{equation}
|\mathcal{S}_{x}^{\left(n\right)}|=0,\quad
|\mathcal{S}_{y}^{\left(n\right)}|=\begin{cases}
0 & p<\frac{1}{2}\\
\left[\frac{2}{3}\left(1-p\right)\right]^{2} & p>\frac{1}{2}
\end{cases}.
\end{equation}
This highlights the distinct behavior between the $(-1,\ -1)$ AKLT phase and the $(-1,\ +1)$ intrinsic weak-symmetry protected phase. Specifically, the analytical results for $\mathcal{S}_x$ and $\mathcal{S}_y$ are
\begin{multline}
\mathcal{S}_{x}=\left[\frac{2}{3}\left(1-p\right)\right]^{2}\left[\left(-\frac{1}{3}\right)^{l}+\left(-\frac{1}{3}\right)^{N-l}\right]\\
\rightarrow \left[\frac{2}{3}\left(1-p\right)\right]^{2}(-\frac{1}{3})^l,
\end{multline}
and
\begin{multline}
\mathcal{S}_y=\left[\frac{2}{3}\left(1-p\right)\right]^{2}\left[\left(\frac{-1+4p}{3}\right)^{l}+\left(-\frac{1}{3}\right)^{N}\right]\\
\rightarrow \left[\frac{2}{3}\left(1-p\right)\right]^{2}\left(\frac{-1+4p}{3}\right)^{l},
\end{multline}
while the leading term of the normalization factor $|\text{Tr}\left(\rho R_{z}^{\otimes N}\right)|^{l/N}$ [see Eq.~\eqref{eq:normalized_string}] is given by
\begin{equation}
|\text{Tr}\left(\rho R_{z}^{\otimes N}\right)|^{l/N}\rightarrow\begin{cases}
\left(\frac{3-4p}{3}\right)^{l} & p<\frac{1}{2}\\
\left(\frac{-1+4p}{3}\right)^{l} & p>\frac{1}{2}
\end{cases}.
\end{equation}

\clearpage
\begin{center}
\textbf{\large Supplemental Material for ``Anomalous matrix product operator symmetries and 1D mixed-state phases''}
\end{center}
\setcounter{table}{0}
\renewcommand{\thetable}{S\arabic{table}}%
\setcounter{figure}{0}
\renewcommand{\thefigure}{S\arabic{figure}}
\setcounter{equation}{0}
\renewcommand{\theequation}{S\arabic{equation}}
\section{Insertion of symmetry flux \label{supp_sec:Symmetry_flux_insertion}}
We demonstrate the insertion of a symmetry flux into the density matrix and its tensor network representation.

\begin{figure}[htbp]
\includegraphics[scale=0.18]{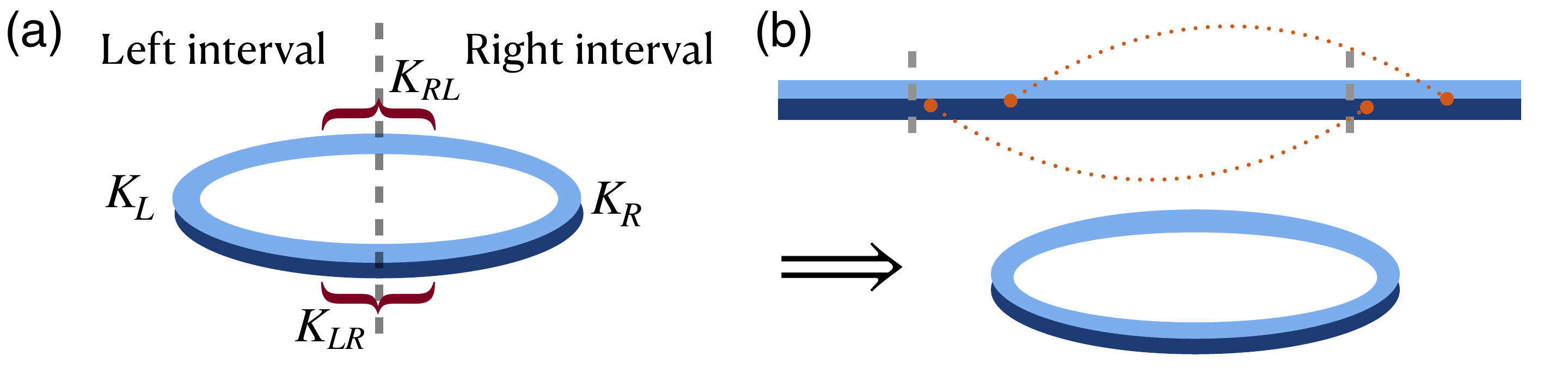}
\caption{Illustration of symmetry flux insertion. The modular Hamiltonian $K$ is decomposed into four terms: $K_L$ and $K_R$, supported on the left and right intervals, together with $K_{LR}$ and $K_{RL}$, which couple them. The twisted boundary condition is implemented through conjugating $K_{LR}$ by $U_{g_1}^{(R)}$,  supported on the right interval.  Panel (b) shows how an $N$-site periodic chain can be obtained from an infinite chain by identifying every $N$ sites. \label{supp_fig:flux_insertion}}
\end{figure}

For a density matrix $\rho=e^{-K}$ with $K$ a Hermitian
operator composed of local terms, we introduce a symmetry flux via a twisted boundary condition~\cite{zaletel2013prl,zaletel2014njp}. Specifically, we consider a one-dimensional ring divided into left and right intervals [see Fig.~\ref{supp_fig:flux_insertion}(a)]. The
modular Hamiltonian with a $g_1$ flux, $K(g_1)$, is given by 
\begin{equation}
K\left(g_1\right)=\left(K_{L}+K_{R}\right)+U_{g_1}^{\left(R\right)}K_{LR}U_{g_1}^{\left(R\right)\dagger}+K_{RL},
\end{equation}
where $K_{L}$ and $K_{R}$ act on the left and right intervals, respectively, while $K_{LR}$ and $K_{RL}$ couple the two intervals. The $g_1$-symmetry flux is implemented through a twisted boundary condition, where we conjugate $K_{LR}$ by $U_{g_1}^{\left(R\right)}$, a unitary that applies $u_{g_1}$ to each site in the right interval and acts trivially elsewhere. 

For a locally purified density operator (LPDO), implementing the symmetry flux (twisted boundary condition) involves treating the one-dimensional $N$-site periodic chain as a unit cell within a translationally invariant infinite chain~\cite{zaletel2013prl}, i.e., as the quotient space $\mathbb{Z}/(N\mathbb{Z})$ [see Fig.~\ref{supp_fig:flux_insertion}(b)]. From this, the LPDO for the periodic chain is derived from the infinite version. The twisted boundary term $U_{g_1}^{\left(R\right)}K_{LR}U_{g_1}^{\left(R\right)\dagger}$
can be mimicked on the infinite chain by applying to the density matrix, for every integer $n$, a local transformation $(u_{g_1})^n$ on each site within segment $\left[nN+1,\left(n+1\right)N\right]$. This yields the flux-inserted LPDO for the periodic chain by taking the $N$-site unit cell:
\begin{equation}
\rho(V_{g_1})=
\begin{array}{c}
	\begin{tikzpicture}[scale=.5, baseline={([yshift=0ex]current bounding box.center)}, thick]
		\draw[shift={(0,0)},dotted] (0,0) -- (6,0);
        \draw[shift={(0,0)},dotted] (0,2) -- (6,2);
        \draw[shift={(0,0)}] (-2.2,0) arc(90:270:0.2);
        \draw[shift={(0,2)}] (-2.2,0) arc(90:270:0.2);
        \draw[shift={(0,0)}] (6.6,-0.4) arc(-90:90:0.2);
        \draw[shift={(0,2)}] (6.6,-0.4) arc(-90:90:0.2);
		\ATensorLPDO{0.4,2}{$A$}
        \ADaggTensorLPDO{0.4,0}{$A^*$}
        \vTensor{-1.9,0}{$V_{g_1}^*$}
		\ATensorLPDO{5,2}{$A$}
        \ADaggTensorLPDO{5,0}{$A^*$}
        \vTensor{-1.9,2}{$V_{g_1}$}
        \draw[red] (1.7,-1.2)--(1.7,3.2);
        \draw[red] (6.3,-1.2)--(6.3,3.2);
        \draw[red] (1.7,3.2) arc(0:180:0.35);
        \draw[red] (6.3,3.2) arc(0:180:0.35);
        \draw[red] (1.7,-1.2) arc(0:-180:0.35);
        \draw[red] (6.3,-1.2) arc(0:-180:0.35);
        \draw[preaction={draw=white, line width=5pt}] (1.5,0) --(2,0);
        \draw[preaction={draw=white, line width=5pt}] (1.5,2) --(2,2);
        \draw[preaction={draw=white, line width=5pt}] (6.1,0) --(6.6,0);
        \draw[preaction={draw=white, line width=5pt}] (6.1,2) --(6.6,2);

	\end{tikzpicture}
	\end{array},
\end{equation}
where $\rho(\mathbb{1})$ (no-flux insertion) is denoted by $\rho$ when not explicitly stated otherwise. 

\section{Proof of the string order parameter selection rule}
We demonstrate the selection rule for the string order parameter, which is reproduced here for convenience, 
\begin{equation}\label{supp_eq:selection_normalized_sop}
\frac{\mathcal{S}(g_{2},\chi_{g_{1}}^{L},\chi_{g_{1}}^{R})}{[\lambda_{g_2}^{(0)}]^{l}}|_{l\rightarrow\infty}=\begin{cases}
\mathcal{O}\left(1\right),\ \text{if}\ e^{i\mathcal{Q}(g_{1},g_{2})}= e^{i\phi(g_{1})}\\
0,\quad\text{if}\ e^{i\mathcal{Q}(g_{1},g_{2})}\neq e^{i\phi(g_{1})}
\end{cases},
\end{equation}
with $\lambda_{g_2}^{(0)}$ the leading eigenvalues of the transfer matrix $T(g_2)$. Also, the string order parameter $\mathcal{S}$ is defined as 
\begin{equation}
\mathcal{S}(g_2,\chi^L_{g_1},\chi^R_{g_1})=\text{Tr}\left[\rho\ \chi^L_{g_1}\otimes\left(\otimes_{i=j}^{j+l-1} u_{g_2} \right)\otimes\chi^R_{g_1}\right],
\end{equation}
with $\chi_{g_1}^{L/R}$ satisfying 
\begin{equation}\label{supp_eq:chi_charge}
u_{g_1}\chi_{g_1}^{L/R} u_{g_1}^\dagger = e^{\pm i \phi(g_1)}\chi_{g_1}^{L/R}.
\end{equation}  
We now recast the string order parameter as
\begin{equation}
\mathcal{S}(g_2,\chi_{g_1}^L,\chi_{g_1}^R)=\text{tr}\left[|B^R_{g_1})(B^L_{g_1}|T(g_2)^l\right],
\end{equation}
where tracing out the complement of the string subsystem yields the boundary vectors $|B_{g_1}^R)$ and $(B_{g_1}^L|$ associated with $\chi_{g_1}^{L/R}$, defined diagrammatically as:
\begin{equation}\label{eq:eigenvalue_eq}
(B_{g_1}^L|\equiv
\begin{tikzpicture}[scale=.5, baseline={([yshift=0ex]current bounding box.center)}, thick]
		\draw[rounded corners] (-1.2,2.8)--(-1.7,2.8)--(-1.7,0)--(-1.2,0);
		\uTensor{-1.7,0.9}{$L_{\mathbb{1}}^0$}
		\ATensorLPDO{0.4,0}{$A$}
        		\ADaggTensorLPDO{0.4,2.8}{$A^*$}
       		 \uTensor{-0.2,0.9} {$\chi^L_{g_1}$}
		  \draw[red](1,0.7)--(1,2);
	\end{tikzpicture},
    \quad |B_{g_1}^R)\equiv
    \begin{tikzpicture}[scale=.5, baseline={([yshift=0ex]current bounding box.center)}, thick]
		\ATensorLPDO{0.4,0}{$A$}
        		\ADaggTensorLPDO{0.4,2.8}{$A^*$}
       		 \uTensor{-0.2,0.9} {$\chi^R_{g_1}$}
		  \draw[red](1,0.7)--(1,2);
		  \draw[rounded corners] (1.8,2.8)--(2.3,2.8)--(2.3,0)--(1.8,0);
		\uTensor{2.3,0.9}{$R_{\mathbb{1}}^0$}
	\end{tikzpicture}.
\end{equation} 
Crucially, the symmetry properties of  $\chi^{L/R}_{g_1}$ ensure that the vector $(B^L_{g_1}|$ is a left eigenvector of $V_{g_1}^{*}\otimes V_{g_1}$ with eigenvalue $e^{i\phi(g_1)}$. This follows from Eq.~\eqref{supp_eq:chi_charge} and the symmetry transformation rule of $A$ given by Eq.~({\color{red}{7}}) of the main text. For convenience, Eq.~({\color{red}{7}}) is reproduced here:
\begin{equation}\label{supp_eq:symmetry_trans}
\begin{array}{c}
\begin{tikzpicture}[scale=.5,thick,baseline={([yshift=-9ex]current bounding box.center)}]
			\ATensorLPDO{0,0}{$A$}
			\uTensor{-0.6,0.9}{$u_g$}
            \uTensorA{0.6,0.9}{$u_g^a$}
\end{tikzpicture}
\end{array}
=e^{i\theta_g}
\begin{array}{c}
\begin{tikzpicture}[scale=.5,thick,baseline={([yshift=-4ex]current bounding box.center)}]
			\ATensorLPDO{0,0}{$A$}
			\vTensor{-2.2,0}{$V_g$}
            \vTensor{1.2,0}{$V_g^\dagger$}
\end{tikzpicture}
\end{array}.
\end{equation}
Similarly, $|B^R_{g_1})$ is the corresponding right eigenvector. Thus, the boundary matrix $|B_{g_1}^L)(B_{g_1}^R|$ projects $T(g_2)$ onto the $e^{i\phi(g_1)}$-eigenspace of $V_{g_1}^*\otimes V_{g_1}$. Together with the identity $[V_{g_1}^*\otimes V_{g_1},T(g_2)]=0$ (proven in Appendix {\color{red}{A}} of the main text), this yields the selection rule Eq.~\eqref{supp_eq:selection_normalized_sop}, since the dominant contribution $\sim\left[\lambda_{g_2}^{(0)}\right]^l$ in $T(g_2)^l$ comes from the leading eigenvector of $T(g_2)$, which has eigenvalue $e^{i\mathcal{Q}(g_1, g_2)}$ under $V_{g_1}^*\otimes V_{g_1}$. 

\section{Review of the average symmetry-protected topological phases: A perspective from quantized response \label{supp_sec:aSPT}}
For completeness, we review the average symmetry-protected topological (ASPT) phase \cite{ma2023prx} in one dimension, which unlike our work, requires strong symmetry. Nevertheless, we demonstrate that its classification  can be reproduced by considering the quantized \textit{strong}-symmetry charge response to \textit{weak}-symmetry fluxes
\begin{equation}
e^{i\mathcal{Q}(w,h)} \equiv \frac{\text{Tr}\left[ \rho(V_w)\ U_h\right]}{\text{Tr}\left( \rho\ U_h\right)},
\end{equation}
where we consider an onsite symmetry group $G=H\times W$, with $H$ and $W$ corresponding to strong and weak symmetries, respectively. Here $U_h$ denotes the symmetry operator corresponding to $h\in H$, and $\rho(V_w)$  (or $\rho$) represents the density matrix with (without) weak symmetry fluxes inserted.
The amplitude $|\text{Tr} \left[\rho(V_w)\ U_h\right]|$ (and similarly for $|\text{Tr} \left(\rho\ U_h\right)|$) equals $1$, because both $\rho(V_w)$ and $\rho$ respect strong symmetry $H$. 
This response $e^{i\mathcal{Q}(w,h)}$ is classified by the group cohomology, 
\begin{equation}
e^{i\mathcal{Q}(w,h)}\in \mathcal{H}^1(W, \mathcal{H}^1(H, U(1))),
\end{equation}
with $\mathcal{H}^1(\dots)$ for the first cohomology group, as $e^{i\mathcal{Q}(w,h)}$ forms a representation for both $W$ and $H$.  Including topological phases protected by solely strong symmetry, captured by $\mathcal{H}^2(H, U(1))$, this classification reproduces the result in Ref.~\cite{ma2023prx}.

\section{Conservation law of topological charge}
We establish an identity that connects the quantized responses of the physical system $e^{i\mathcal{Q}(g_1,g_2)}$, the ancillary system $e^{i\mathcal{Q}_a(g_1,g_2)}$, and the underlying purified wavefunction $e^{i\mathcal{Q}_t(g_1,g_2)}\equiv \frac{\omega(g_1,g_2)}{\omega(g_2,g_1)}$:  
\begin{equation}
e^{i\mathcal{Q}_t(g_1,g_2)}= e^{i\mathcal{Q}(g_1,g_2)}\times e^{i\mathcal{Q}_a(g_1,g_2)}.\label{supp_eq:topological_charge_conservation}
\end{equation}
This represents a conservation law of topological charge,
\begin{equation}
\mathcal{Q}_t(g_1,g_2) =\mathcal{Q}(g_1,g_2)+\mathcal{Q}_a(g_1,g_2),
\end{equation}
stating that the total topological charge $\mathcal{Q}_t(g_1,g_2)$ is the sum of the charges from the physical system $\mathcal{Q}(g_1,g_2)$ and the ancillary system $\mathcal{Q}_a(g_1,g_2)$. 
In the case of pure states and ASPT phases (as discussed in the previous section), the ancillary system is charge-free [i.e., $\mathcal{Q}_a(g_1,g_2)=0$] due to $g_2$ being a strong symmetry, yielding 
\begin{equation}
\mathcal{Q}_t(g_1,g_2) =\mathcal{Q}(g_1,g_2),
\end{equation}
classified by the group cohomology. By contrast, in our case, when both $g_1$ and $g_2$ are weak symmetries, $\mathcal{Q}_a(g_1,g_2)$ can be non-zero, signaling intrinsic mixed-state phases protected by weak symmetries, thereby extending the existing framework.

The derivation of Eq.~\eqref{supp_eq:topological_charge_conservation} follows from Eq.~\eqref{supp_eq:symmetry_trans}, the symmetry transformation rule of tensor $A$. 
The tensor-network representation of $e^{i\mathcal{Q}(g_1,g_2)}$ is
\begin{equation}
e^{i\mathcal{Q}(g_1,g_2)}=
\frac{1}{\text{Tr}\left(\rho  U_{g_2}\right)}
\begin{array}{c}
	\begin{tikzpicture}[scale=.5, baseline={([yshift=0ex]current bounding box.center)}, thick]
		\draw[shift={(0,0)},dotted] (0,0) -- (6,0);
        \draw[shift={(0,0)},dotted] (0,2.8) -- (6,2.8);
        \draw[shift={(0,0)}] (-2.2,0) arc(90:270:0.2);
        \draw[shift={(0,2.8)}] (-2.2,0) arc(90:270:0.2);
        \draw[shift={(0,0)}] (6.6,-0.4) arc(-90:90:0.2);
        \draw[shift={(0,0)}] (6.6,2.4) arc(-90:90:0.2);
		\ATensorLPDO{0.4,0}{$A$}
        \ADaggTensorLPDO{0.4,2.8}{$A^*$}
        \uTensor{-0.2,0.9}
        {$u_{g_2}$}
        \vTensor{-1.9,0}{$V_{g_1}$}
		\ATensorLPDO{5,0}{$A$}
        \ADaggTensorLPDO{5,2.8}{$A^*$}
        \uTensor{4.4,0.9}{$u_{g_2}$}
        \vTensor{-1.9,2.8}{$V_{g_1}^*$}
        \draw[red](1,0.7)--(1,2);
        \draw[red](5.6,0.7)--(5.6,2);
	\end{tikzpicture}
	\end{array},
\end{equation}
which via Eq.~\eqref{supp_eq:symmetry_trans}, becomes
\begin{eqnarray}
e^{i\mathcal{Q}(g_1,g_2)}&=&\frac{\omega(g_1,g_2)}{\omega(g_2,g_1)}\nonumber\\
{}&&
\times \frac{e^{iN\theta_{g_2}}}{\text{Tr}\left(\rho U_{g_2}\right)}
\begin{array}{c}
	\begin{tikzpicture}[scale=.5, baseline={([yshift=0ex]current bounding box.center)}, thick]
		\draw[shift={(0,0)},dotted] (0,0) -- (6,0);
        \draw[shift={(0,0)},dotted] (0,2.8) -- (6,2.8);
        
        \draw[shift={(0,0)}] (-2.2,0) arc(90:270:0.2);
        \draw[shift={(0,2.8)}] (-2.2,0) arc(90:270:0.2);
        
        \draw[shift={(0,0)}] (6.6,-0.4) arc(-90:90:0.2);
        \draw[shift={(0,0)}] (6.6,2.4) arc(-90:90:0.2);
        
		\ATensorLPDO{0.4,0}{$A$}
        \ADaggTensorLPDO{0.4,2.8}{$A^*$}
        
        \uTensorA{1,0.9}{$u_{g_2}^{a\dagger}$}
        
        \vTensor{-1.9,0}{$V_{g_1}$}
        
		\ATensorLPDO{5,0}{$A$}
        \ADaggTensorLPDO{5,2.8}{$A^*$}
        
        \uTensorA{5.6,0.9}{$u_{g_2}^{a\dagger}$}
        \vTensor{-1.9,2.8}{$V_{g_1}^*$}
        \draw[red](1,0.7)--(1,0.8);
        \draw[red](5.6,0.7)--(5.6,0.8);

        \draw[black](-0.2,0.6)--(-0.2,2);
        \draw[black](4.4,0.6)--(4.4,2);
	\end{tikzpicture}
	\end{array}.\nonumber\\
\end{eqnarray}
This gives the relation between $e^{i\mathcal{Q}(g_1,g_2)}$ and $e^{i\mathcal{Q}_a(g_1,g_2)}$,
\begin{equation}
e^{i\mathcal{Q}(g_1,g_2)}= \frac{\omega(g_1,g_2)}{\omega(g_2,g_1)}\times e^{-i\mathcal{Q}_a(g_1, g_2)}.
\end{equation}
Here we have used $\frac{e^{iN\theta_{g_2}}}{\text{Tr}\left(\rho U_{g_2}\right)}=\frac{1}{\text{Tr}_a\left( \rho_a U_{g_2}^{a\dagger}\right)}$, where $\rho_a$ is the density matrix of the ancillas, obtained by tracing out the physical system from the combined pure state, and $U_{g_2}^a=(u_{g_2}^a)^{\otimes N}$ denotes the symmetry action on the ancillas. The tensor-network representation of $e^{i\mathcal{Q}_a(g_1,g_2)}$ is
\begin{equation}
e^{i\mathcal{Q}_a(g_1,g_2)}=\frac{1}{\text{Tr}_a\left(\rho_a U^a_{g_2}\right)}
\begin{array}{c}
\begin{tikzpicture}[scale=.5, baseline={([yshift=0ex]current bounding box.center)}, thick]
		\draw[shift={(0,0)},dotted] (0,0) -- (6,0);
        \draw[shift={(0,0)},dotted] (0,2.8) -- (6,2.8);
        
        \draw[shift={(0,0)}] (-2.2,0) arc(90:270:0.2);
        \draw[shift={(0,2.8)}] (-2.2,0) arc(90:270:0.2);
        
        \draw[shift={(0,0)}] (6.6,-0.4) arc(-90:90:0.2);
        \draw[shift={(0,0)}] (6.6,2.4) arc(-90:90:0.2);
        
		\ATensorLPDO{0.4,0}{$A$}
        \ADaggTensorLPDO{0.4,2.8}{$A^*$}
        
        \uTensorA{1,0.9}{$u_{g_2}^{a}$}
        
        \vTensor{-1.9,0}{$V_{g_1}$}
        
		\ATensorLPDO{5,0}{$A$}
        \ADaggTensorLPDO{5,2.8}{$A^*$}
        
        \uTensorA{5.6,0.9}{$u_{g_2}^{a}$}
        \vTensor{-1.9,2.8}{$V_{g_1}^*$}
        \draw[red](1,0.7)--(1,0.8);
        \draw[red](5.6,0.7)--(5.6,0.8);

        \draw[black](-0.2,0.6)--(-0.2,2);
        \draw[black](4.4,0.6)--(4.4,2);
	\end{tikzpicture}
	\end{array}.
\end{equation}
This completes the proof of Eq.~\eqref{supp_eq:topological_charge_conservation}.

\begin{thebibliography}{64}%
\makeatletter
\providecommand \@ifxundefined [1]{%
 \@ifx{#1\undefined}
}%
\providecommand \@ifnum [1]{%
 \ifnum #1\expandafter \@firstoftwo
 \else \expandafter \@secondoftwo
 \fi
}%
\providecommand \@ifx [1]{%
 \ifx #1\expandafter \@firstoftwo
 \else \expandafter \@secondoftwo
 \fi
}%
\providecommand \natexlab [1]{#1}%
\providecommand \enquote  [1]{``#1''}%
\providecommand \bibnamefont  [1]{#1}%
\providecommand \bibfnamefont [1]{#1}%
\providecommand \citenamefont [1]{#1}%
\providecommand \href@noop [0]{\@secondoftwo}%
\providecommand \href [0]{\begingroup \@sanitize@url \@href}%
\providecommand \@href[1]{\@@startlink{#1}\@@href}%
\providecommand \@@href[1]{\endgroup#1\@@endlink}%
\providecommand \@sanitize@url [0]{\catcode `\\12\catcode `\$12\catcode
  `\&12\catcode `\#12\catcode `\^12\catcode `\_12\catcode `\%12\relax}%
\providecommand \@@startlink[1]{}%
\providecommand \@@endlink[0]{}%
\providecommand \url  [0]{\begingroup\@sanitize@url \@url }%
\providecommand \@url [1]{\endgroup\@href {#1}{\urlprefix }}%
\providecommand \urlprefix  [0]{URL }%
\providecommand \Eprint [0]{\href }%
\providecommand \doibase [0]{http://dx.doi.org/}%
\providecommand \selectlanguage [0]{\@gobble}%
\providecommand \bibinfo  [0]{\@secondoftwo}%
\providecommand \bibfield  [0]{\@secondoftwo}%
\providecommand \translation [1]{[#1]}%
\providecommand \BibitemOpen [0]{}%
\providecommand \bibitemStop [0]{}%
\providecommand \bibitemNoStop [0]{.\EOS\space}%
\providecommand \EOS [0]{\spacefactor3000\relax}%
\providecommand \BibitemShut  [1]{\csname bibitem#1\endcsname}%
\let\auto@bib@innerbib\@empty
\bibitem [{\citenamefont {Bu{\v{c}}a}\ and\ \citenamefont
  {Prosen}(2012)}]{buca2012njp}%
  \BibitemOpen
  \bibfield  {author} {\bibinfo {author} {\bibfnamefont {B.}~\bibnamefont
  {Bu{\v{c}}a}}\ and\ \bibinfo {author} {\bibfnamefont {T.}~\bibnamefont
  {Prosen}},\ }\href {\doibase 10.1088/1367-2630/14/7/073007} {\bibfield
  {journal} {\bibinfo  {journal} {New Journal of Physics}\ }\textbf {\bibinfo
  {volume} {14}},\ \bibinfo {pages} {073007} (\bibinfo {year}
  {2012})}\BibitemShut {NoStop}%
\bibitem [{\citenamefont {Viyuela}\ \emph {et~al.}(2014)\citenamefont
  {Viyuela}, \citenamefont {Rivas},\ and\ \citenamefont
  {Martin-Delgado}}]{viyuela2014prl1d}%
  \BibitemOpen
  \bibfield  {author} {\bibinfo {author} {\bibfnamefont {O.}~\bibnamefont
  {Viyuela}}, \bibinfo {author} {\bibfnamefont {A.}~\bibnamefont {Rivas}}, \
  and\ \bibinfo {author} {\bibfnamefont {M.~A.}\ \bibnamefont
  {Martin-Delgado}},\ }\href {\doibase 10.1103/PhysRevLett.112.130401}
  {\bibfield  {journal} {\bibinfo  {journal} {Phys. Rev. Lett.}\ }\textbf
  {\bibinfo {volume} {112}},\ \bibinfo {pages} {130401} (\bibinfo {year}
  {2014})}\BibitemShut {NoStop}%
\bibitem [{\citenamefont {Budich}\ and\ \citenamefont
  {Diehl}(2015)}]{budich2015prb}%
  \BibitemOpen
  \bibfield  {author} {\bibinfo {author} {\bibfnamefont {J.~C.}\ \bibnamefont
  {Budich}}\ and\ \bibinfo {author} {\bibfnamefont {S.}~\bibnamefont {Diehl}},\
  }\href {\doibase 10.1103/PhysRevB.91.165140} {\bibfield  {journal} {\bibinfo
  {journal} {Phys. Rev. B}\ }\textbf {\bibinfo {volume} {91}},\ \bibinfo
  {pages} {165140} (\bibinfo {year} {2015})}\BibitemShut {NoStop}%
\bibitem [{\citenamefont {Bardyn}\ \emph {et~al.}(2018)\citenamefont {Bardyn},
  \citenamefont {Wawer}, \citenamefont {Altland}, \citenamefont
  {Fleischhauer},\ and\ \citenamefont {Diehl}}]{bardyn2018prx}%
  \BibitemOpen
  \bibfield  {author} {\bibinfo {author} {\bibfnamefont {C.-E.}\ \bibnamefont
  {Bardyn}}, \bibinfo {author} {\bibfnamefont {L.}~\bibnamefont {Wawer}},
  \bibinfo {author} {\bibfnamefont {A.}~\bibnamefont {Altland}}, \bibinfo
  {author} {\bibfnamefont {M.}~\bibnamefont {Fleischhauer}}, \ and\ \bibinfo
  {author} {\bibfnamefont {S.}~\bibnamefont {Diehl}},\ }\href {\doibase
  10.1103/PhysRevX.8.011035} {\bibfield  {journal} {\bibinfo  {journal} {Phys.
  Rev. X}\ }\textbf {\bibinfo {volume} {8}},\ \bibinfo {pages} {011035}
  (\bibinfo {year} {2018})}\BibitemShut {NoStop}%
\bibitem [{\citenamefont {Huang}\ \emph {et~al.}(2022)\citenamefont {Huang},
  \citenamefont {Sun},\ and\ \citenamefont {Diehl}}]{huang2022prb}%
  \BibitemOpen
  \bibfield  {author} {\bibinfo {author} {\bibfnamefont {Z.-M.}\ \bibnamefont
  {Huang}}, \bibinfo {author} {\bibfnamefont {X.-Q.}\ \bibnamefont {Sun}}, \
  and\ \bibinfo {author} {\bibfnamefont {S.}~\bibnamefont {Diehl}},\ }\href
  {\doibase 10.1103/PhysRevB.106.245204} {\bibfield  {journal} {\bibinfo
  {journal} {Phys. Rev. B}\ }\textbf {\bibinfo {volume} {106}},\ \bibinfo
  {pages} {245204} (\bibinfo {year} {2022})}\BibitemShut {NoStop}%
\bibitem [{\citenamefont {Altland}\ \emph {et~al.}(2021)\citenamefont
  {Altland}, \citenamefont {Fleischhauer},\ and\ \citenamefont
  {Diehl}}]{altland2021prx}%
  \BibitemOpen
  \bibfield  {author} {\bibinfo {author} {\bibfnamefont {A.}~\bibnamefont
  {Altland}}, \bibinfo {author} {\bibfnamefont {M.}~\bibnamefont
  {Fleischhauer}}, \ and\ \bibinfo {author} {\bibfnamefont {S.}~\bibnamefont
  {Diehl}},\ }\href {\doibase 10.1103/PhysRevX.11.021037} {\bibfield  {journal}
  {\bibinfo  {journal} {Phys. Rev. X}\ }\textbf {\bibinfo {volume} {11}},\
  \bibinfo {pages} {021037} (\bibinfo {year} {2021})}\BibitemShut {NoStop}%
\bibitem [{\citenamefont {Huang}\ and\ \citenamefont
  {Diehl}(2024)}]{huang2024arxiv_mstop}%
  \BibitemOpen
  \bibfield  {author} {\bibinfo {author} {\bibfnamefont {Z.-M.}\ \bibnamefont
  {Huang}}\ and\ \bibinfo {author} {\bibfnamefont {S.}~\bibnamefont {Diehl}},\
  }\href@noop {} {\enquote {\bibinfo {title} {Mixed state topological order
  parameters for symmetry protected fermion matter},}\ } (\bibinfo {year}
  {2024}),\ \Eprint {http://arxiv.org/abs/2401.10993} {arXiv:2401.10993
  [cond-mat.quant-gas]} \BibitemShut {NoStop}%
\bibitem [{\citenamefont {Mao}\ \emph {et~al.}(2024{\natexlab{a}})\citenamefont
  {Mao}, \citenamefont {Yang},\ and\ \citenamefont {Zhai}}]{mao2024rpp}%
  \BibitemOpen
  \bibfield  {author} {\bibinfo {author} {\bibfnamefont {L.}~\bibnamefont
  {Mao}}, \bibinfo {author} {\bibfnamefont {F.}~\bibnamefont {Yang}}, \ and\
  \bibinfo {author} {\bibfnamefont {H.}~\bibnamefont {Zhai}},\ }\href {\doibase
  10.1088/1361-6633/ad44d4} {\bibfield  {journal} {\bibinfo  {journal} {Reports
  on Progress in Physics}\ }\textbf {\bibinfo {volume} {87}},\ \bibinfo {pages}
  {070501} (\bibinfo {year} {2024}{\natexlab{a}})}\BibitemShut {NoStop}%
\bibitem [{\citenamefont {Huang}\ and\ \citenamefont
  {Diehl}(2025)}]{huang2025prl}%
  \BibitemOpen
  \bibfield  {author} {\bibinfo {author} {\bibfnamefont {Z.-M.}\ \bibnamefont
  {Huang}}\ and\ \bibinfo {author} {\bibfnamefont {S.}~\bibnamefont {Diehl}},\
  }\href {\doibase 10.1103/PhysRevLett.134.053002} {\bibfield  {journal}
  {\bibinfo  {journal} {Phys. Rev. Lett.}\ }\textbf {\bibinfo {volume} {134}},\
  \bibinfo {pages} {053002} (\bibinfo {year} {2025})}\BibitemShut {NoStop}%
\bibitem [{\citenamefont {Ma}\ and\ \citenamefont {Wang}(2023)}]{ma2023prx}%
  \BibitemOpen
  \bibfield  {author} {\bibinfo {author} {\bibfnamefont {R.}~\bibnamefont
  {Ma}}\ and\ \bibinfo {author} {\bibfnamefont {C.}~\bibnamefont {Wang}},\
  }\href {\doibase 10.1103/PhysRevX.13.031016} {\bibfield  {journal} {\bibinfo
  {journal} {Phys. Rev. X}\ }\textbf {\bibinfo {volume} {13}},\ \bibinfo
  {pages} {031016} (\bibinfo {year} {2023})}\BibitemShut {NoStop}%
\bibitem [{\citenamefont {Coser}\ and\ \citenamefont
  {P{\'{e}}rez-Garc{\'{i}}a}(2019)}]{coser2019quantum}%
  \BibitemOpen
  \bibfield  {author} {\bibinfo {author} {\bibfnamefont {A.}~\bibnamefont
  {Coser}}\ and\ \bibinfo {author} {\bibfnamefont {D.}~\bibnamefont
  {P{\'{e}}rez-Garc{\'{i}}a}},\ }\href {\doibase 10.22331/q-2019-08-12-174}
  {\bibfield  {journal} {\bibinfo  {journal} {{Quantum}}\ }\textbf {\bibinfo
  {volume} {3}},\ \bibinfo {pages} {174} (\bibinfo {year} {2019})}\BibitemShut
  {NoStop}%
\bibitem [{\citenamefont {de~Groot}\ \emph {et~al.}(2022)\citenamefont
  {de~Groot}, \citenamefont {Turzillo},\ and\ \citenamefont
  {Schuch}}]{degroot2022quantum}%
  \BibitemOpen
  \bibfield  {author} {\bibinfo {author} {\bibfnamefont {C.}~\bibnamefont
  {de~Groot}}, \bibinfo {author} {\bibfnamefont {A.}~\bibnamefont {Turzillo}},
  \ and\ \bibinfo {author} {\bibfnamefont {N.}~\bibnamefont {Schuch}},\ }\href
  {\doibase 10.22331/q-2022-11-10-856} {\bibfield  {journal} {\bibinfo
  {journal} {{Quantum}}\ }\textbf {\bibinfo {volume} {6}},\ \bibinfo {pages}
  {856} (\bibinfo {year} {2022})}\BibitemShut {NoStop}%
\bibitem [{\citenamefont {Guo}\ \emph {et~al.}(2024)\citenamefont {Guo},
  \citenamefont {Zhang}, \citenamefont {Zhang}, \citenamefont {Yang},\ and\
  \citenamefont {Bi}}]{guo2024arxiv_lpdo}%
  \BibitemOpen
  \bibfield  {author} {\bibinfo {author} {\bibfnamefont {Y.}~\bibnamefont
  {Guo}}, \bibinfo {author} {\bibfnamefont {J.-H.}\ \bibnamefont {Zhang}},
  \bibinfo {author} {\bibfnamefont {H.-R.}\ \bibnamefont {Zhang}}, \bibinfo
  {author} {\bibfnamefont {S.}~\bibnamefont {Yang}}, \ and\ \bibinfo {author}
  {\bibfnamefont {Z.}~\bibnamefont {Bi}},\ }\href@noop {} {\enquote {\bibinfo
  {title} {Locally purified density operators for symmetry-protected
  topological phases in mixed states},}\ } (\bibinfo {year} {2024}),\ \Eprint
  {http://arxiv.org/abs/2403.16978} {arXiv:2403.16978 [cond-mat.str-el]}
  \BibitemShut {NoStop}%
\bibitem [{\citenamefont {Xue}\ \emph {et~al.}(2024)\citenamefont {Xue},
  \citenamefont {Lee},\ and\ \citenamefont {Bao}}]{xue2024arxiv_mpdo}%
  \BibitemOpen
  \bibfield  {author} {\bibinfo {author} {\bibfnamefont {H.}~\bibnamefont
  {Xue}}, \bibinfo {author} {\bibfnamefont {J.~Y.}\ \bibnamefont {Lee}}, \ and\
  \bibinfo {author} {\bibfnamefont {Y.}~\bibnamefont {Bao}},\ }\href@noop {}
  {\enquote {\bibinfo {title} {Tensor network formulation of symmetry protected
  topological phases in mixed states},}\ } (\bibinfo {year} {2024}),\ \Eprint
  {http://arxiv.org/abs/2403.17069} {arXiv:2403.17069 [cond-mat.str-el]}
  \BibitemShut {NoStop}%
\bibitem [{\citenamefont {Kawabata}\ \emph {et~al.}(2024)\citenamefont
  {Kawabata}, \citenamefont {Sohal},\ and\ \citenamefont
  {Ryu}}]{kawabata2024prl}%
  \BibitemOpen
  \bibfield  {author} {\bibinfo {author} {\bibfnamefont {K.}~\bibnamefont
  {Kawabata}}, \bibinfo {author} {\bibfnamefont {R.}~\bibnamefont {Sohal}}, \
  and\ \bibinfo {author} {\bibfnamefont {S.}~\bibnamefont {Ryu}},\ }\href
  {\doibase 10.1103/PhysRevLett.132.070402} {\bibfield  {journal} {\bibinfo
  {journal} {Phys. Rev. Lett.}\ }\textbf {\bibinfo {volume} {132}},\ \bibinfo
  {pages} {070402} (\bibinfo {year} {2024})}\BibitemShut {NoStop}%
\bibitem [{\citenamefont {Lee}\ \emph {et~al.}(2025)\citenamefont {Lee},
  \citenamefont {You},\ and\ \citenamefont {Xu}}]{lee2025quantum}%
  \BibitemOpen
  \bibfield  {author} {\bibinfo {author} {\bibfnamefont {J.~Y.}\ \bibnamefont
  {Lee}}, \bibinfo {author} {\bibfnamefont {Y.-Z.}\ \bibnamefont {You}}, \ and\
  \bibinfo {author} {\bibfnamefont {C.}~\bibnamefont {Xu}},\ }\href {\doibase
  10.22331/q-2025-01-23-1607} {\bibfield  {journal} {\bibinfo  {journal}
  {{Quantum}}\ }\textbf {\bibinfo {volume} {9}},\ \bibinfo {pages} {1607}
  (\bibinfo {year} {2025})}\BibitemShut {NoStop}%
\bibitem [{\citenamefont {Blatt}\ and\ \citenamefont
  {Roos}(2012)}]{blatt2012nature}%
  \BibitemOpen
  \bibfield  {author} {\bibinfo {author} {\bibfnamefont {R.}~\bibnamefont
  {Blatt}}\ and\ \bibinfo {author} {\bibfnamefont {C.~F.}\ \bibnamefont
  {Roos}},\ }\href {\doibase 10.1038/nphys2252} {\bibfield  {journal} {\bibinfo
   {journal} {Nature Physics}\ }\textbf {\bibinfo {volume} {8}},\ \bibinfo
  {pages} {277} (\bibinfo {year} {2012})}\BibitemShut {NoStop}%
\bibitem [{\citenamefont {Monroe}\ \emph {et~al.}(2021)\citenamefont {Monroe},
  \citenamefont {Campbell}, \citenamefont {Duan}, \citenamefont {Gong},
  \citenamefont {Gorshkov}, \citenamefont {Hess}, \citenamefont {Islam},
  \citenamefont {Kim}, \citenamefont {Linke}, \citenamefont {Pagano},
  \citenamefont {Richerme}, \citenamefont {Senko},\ and\ \citenamefont
  {Yao}}]{monroe2021rmp}%
  \BibitemOpen
  \bibfield  {author} {\bibinfo {author} {\bibfnamefont {C.}~\bibnamefont
  {Monroe}}, \bibinfo {author} {\bibfnamefont {W.~C.}\ \bibnamefont
  {Campbell}}, \bibinfo {author} {\bibfnamefont {L.-M.}\ \bibnamefont {Duan}},
  \bibinfo {author} {\bibfnamefont {Z.-X.}\ \bibnamefont {Gong}}, \bibinfo
  {author} {\bibfnamefont {A.~V.}\ \bibnamefont {Gorshkov}}, \bibinfo {author}
  {\bibfnamefont {P.~W.}\ \bibnamefont {Hess}}, \bibinfo {author}
  {\bibfnamefont {R.}~\bibnamefont {Islam}}, \bibinfo {author} {\bibfnamefont
  {K.}~\bibnamefont {Kim}}, \bibinfo {author} {\bibfnamefont {N.~M.}\
  \bibnamefont {Linke}}, \bibinfo {author} {\bibfnamefont {G.}~\bibnamefont
  {Pagano}}, \bibinfo {author} {\bibfnamefont {P.}~\bibnamefont {Richerme}},
  \bibinfo {author} {\bibfnamefont {C.}~\bibnamefont {Senko}}, \ and\ \bibinfo
  {author} {\bibfnamefont {N.~Y.}\ \bibnamefont {Yao}},\ }\href {\doibase
  10.1103/RevModPhys.93.025001} {\bibfield  {journal} {\bibinfo  {journal}
  {Rev. Mod. Phys.}\ }\textbf {\bibinfo {volume} {93}},\ \bibinfo {pages}
  {025001} (\bibinfo {year} {2021})}\BibitemShut {NoStop}%
\bibitem [{\citenamefont {Noel}\ \emph {et~al.}(2022)\citenamefont {Noel},
  \citenamefont {Niroula}, \citenamefont {Zhu}, \citenamefont {Risinger},
  \citenamefont {Egan}, \citenamefont {Biswas}, \citenamefont {Cetina},
  \citenamefont {Gorshkov}, \citenamefont {Gullans}, \citenamefont {Huse},\
  and\ \citenamefont {Monroe}}]{noel2022np}%
  \BibitemOpen
  \bibfield  {author} {\bibinfo {author} {\bibfnamefont {C.}~\bibnamefont
  {Noel}}, \bibinfo {author} {\bibfnamefont {P.}~\bibnamefont {Niroula}},
  \bibinfo {author} {\bibfnamefont {D.}~\bibnamefont {Zhu}}, \bibinfo {author}
  {\bibfnamefont {A.}~\bibnamefont {Risinger}}, \bibinfo {author}
  {\bibfnamefont {L.}~\bibnamefont {Egan}}, \bibinfo {author} {\bibfnamefont
  {D.}~\bibnamefont {Biswas}}, \bibinfo {author} {\bibfnamefont
  {M.}~\bibnamefont {Cetina}}, \bibinfo {author} {\bibfnamefont {A.~V.}\
  \bibnamefont {Gorshkov}}, \bibinfo {author} {\bibfnamefont {M.~J.}\
  \bibnamefont {Gullans}}, \bibinfo {author} {\bibfnamefont {D.~A.}\
  \bibnamefont {Huse}}, \ and\ \bibinfo {author} {\bibfnamefont
  {C.}~\bibnamefont {Monroe}},\ }\href {\doibase 10.1038/s41567-022-01619-7}
  {\bibfield  {journal} {\bibinfo  {journal} {Nature Physics}\ }\textbf
  {\bibinfo {volume} {18}},\ \bibinfo {pages} {760} (\bibinfo {year}
  {2022})}\BibitemShut {NoStop}%
\bibitem [{\citenamefont {Semeghini}\ \emph {et~al.}(2021)\citenamefont
  {Semeghini}, \citenamefont {Levine}, \citenamefont {Keesling}, \citenamefont
  {Ebadi}, \citenamefont {Wang}, \citenamefont {Bluvstein}, \citenamefont
  {Verresen}, \citenamefont {Pichler}, \citenamefont {Kalinowski},
  \citenamefont {Samajdar}, \citenamefont {Omran}, \citenamefont {Sachdev},
  \citenamefont {Vishwanath}, \citenamefont {Greiner}, \citenamefont
  {Vuleti{\'c}},\ and\ \citenamefont {Lukin}}]{semeghini2021science}%
  \BibitemOpen
  \bibfield  {author} {\bibinfo {author} {\bibfnamefont {G.}~\bibnamefont
  {Semeghini}}, \bibinfo {author} {\bibfnamefont {H.}~\bibnamefont {Levine}},
  \bibinfo {author} {\bibfnamefont {A.}~\bibnamefont {Keesling}}, \bibinfo
  {author} {\bibfnamefont {S.}~\bibnamefont {Ebadi}}, \bibinfo {author}
  {\bibfnamefont {T.~T.}\ \bibnamefont {Wang}}, \bibinfo {author}
  {\bibfnamefont {D.}~\bibnamefont {Bluvstein}}, \bibinfo {author}
  {\bibfnamefont {R.}~\bibnamefont {Verresen}}, \bibinfo {author}
  {\bibfnamefont {H.}~\bibnamefont {Pichler}}, \bibinfo {author} {\bibfnamefont
  {M.}~\bibnamefont {Kalinowski}}, \bibinfo {author} {\bibfnamefont
  {R.}~\bibnamefont {Samajdar}}, \bibinfo {author} {\bibfnamefont
  {A.}~\bibnamefont {Omran}}, \bibinfo {author} {\bibfnamefont
  {S.}~\bibnamefont {Sachdev}}, \bibinfo {author} {\bibfnamefont
  {A.}~\bibnamefont {Vishwanath}}, \bibinfo {author} {\bibfnamefont
  {M.}~\bibnamefont {Greiner}}, \bibinfo {author} {\bibfnamefont
  {V.}~\bibnamefont {Vuleti{\'c}}}, \ and\ \bibinfo {author} {\bibfnamefont
  {M.~D.}\ \bibnamefont {Lukin}},\ }\href {\doibase 10.1126/science.abi8794}
  {\bibfield  {journal} {\bibinfo  {journal} {Science}\ }\textbf {\bibinfo
  {volume} {374}},\ \bibinfo {pages} {1242} (\bibinfo {year}
  {2021})}\BibitemShut {NoStop}%
\bibitem [{\citenamefont {Satzinger}\ \emph {et~al.}(2021)\citenamefont
  {Satzinger}, \citenamefont {Liu}, \citenamefont {Smith}, \citenamefont
  {Knapp}, \citenamefont {Newman}, \citenamefont {Jones}, \citenamefont {Chen},
  \citenamefont {Quintana}, \citenamefont {Mi}, \citenamefont {Dunsworth},
  \citenamefont {Gidney}, \citenamefont {Aleiner}, \citenamefont {Arute},
  \citenamefont {Arya}, \citenamefont {Atalaya}, \citenamefont {Babbush} \emph
  {et~al.}}]{satzinger2021science}%
  \BibitemOpen
  \bibfield  {author} {\bibinfo {author} {\bibfnamefont {K.~J.}\ \bibnamefont
  {Satzinger}}, \bibinfo {author} {\bibfnamefont {Y.-J.}\ \bibnamefont {Liu}},
  \bibinfo {author} {\bibfnamefont {A.}~\bibnamefont {Smith}}, \bibinfo
  {author} {\bibfnamefont {C.}~\bibnamefont {Knapp}}, \bibinfo {author}
  {\bibfnamefont {M.}~\bibnamefont {Newman}}, \bibinfo {author} {\bibfnamefont
  {C.}~\bibnamefont {Jones}}, \bibinfo {author} {\bibfnamefont
  {Z.}~\bibnamefont {Chen}}, \bibinfo {author} {\bibfnamefont {C.}~\bibnamefont
  {Quintana}}, \bibinfo {author} {\bibfnamefont {X.}~\bibnamefont {Mi}},
  \bibinfo {author} {\bibfnamefont {A.}~\bibnamefont {Dunsworth}}, \bibinfo
  {author} {\bibfnamefont {C.}~\bibnamefont {Gidney}}, \bibinfo {author}
  {\bibfnamefont {I.}~\bibnamefont {Aleiner}}, \bibinfo {author} {\bibfnamefont
  {F.}~\bibnamefont {Arute}}, \bibinfo {author} {\bibfnamefont
  {K.}~\bibnamefont {Arya}}, \bibinfo {author} {\bibfnamefont {J.}~\bibnamefont
  {Atalaya}}, \bibinfo {author} {\bibnamefont {Babbush}},  \emph {et~al.},\
  }\href {\doibase 10.1126/science.abi8378} {\bibfield  {journal} {\bibinfo
  {journal} {Science}\ }\textbf {\bibinfo {volume} {374}},\ \bibinfo {pages}
  {1237} (\bibinfo {year} {2021})}\BibitemShut {NoStop}%
\bibitem [{\citenamefont {Gross}\ and\ \citenamefont
  {Bloch}(2017)}]{gross2017science}%
  \BibitemOpen
  \bibfield  {author} {\bibinfo {author} {\bibfnamefont {C.}~\bibnamefont
  {Gross}}\ and\ \bibinfo {author} {\bibfnamefont {I.}~\bibnamefont {Bloch}},\
  }\href {\doibase 10.1126/science.aal3837} {\bibfield  {journal} {\bibinfo
  {journal} {Science}\ }\textbf {\bibinfo {volume} {357}},\ \bibinfo {pages}
  {995} (\bibinfo {year} {2017})}\BibitemShut {NoStop}%
\bibitem [{\citenamefont {Ebadi}\ \emph {et~al.}(2021)\citenamefont {Ebadi},
  \citenamefont {Wang}, \citenamefont {Levine}, \citenamefont {Keesling},
  \citenamefont {Semeghini}, \citenamefont {Omran}, \citenamefont {Bluvstein},
  \citenamefont {Samajdar}, \citenamefont {Pichler}, \citenamefont {Ho},
  \citenamefont {Choi}, \citenamefont {Sachdev}, \citenamefont {Greiner},
  \citenamefont {Vuleti{\'c}},\ and\ \citenamefont {Lukin}}]{ebadi2021nature}%
  \BibitemOpen
  \bibfield  {author} {\bibinfo {author} {\bibfnamefont {S.}~\bibnamefont
  {Ebadi}}, \bibinfo {author} {\bibfnamefont {T.~T.}\ \bibnamefont {Wang}},
  \bibinfo {author} {\bibfnamefont {H.}~\bibnamefont {Levine}}, \bibinfo
  {author} {\bibfnamefont {A.}~\bibnamefont {Keesling}}, \bibinfo {author}
  {\bibfnamefont {G.}~\bibnamefont {Semeghini}}, \bibinfo {author}
  {\bibfnamefont {A.}~\bibnamefont {Omran}}, \bibinfo {author} {\bibfnamefont
  {D.}~\bibnamefont {Bluvstein}}, \bibinfo {author} {\bibfnamefont
  {R.}~\bibnamefont {Samajdar}}, \bibinfo {author} {\bibfnamefont
  {H.}~\bibnamefont {Pichler}}, \bibinfo {author} {\bibfnamefont {W.~W.}\
  \bibnamefont {Ho}}, \bibinfo {author} {\bibfnamefont {S.}~\bibnamefont
  {Choi}}, \bibinfo {author} {\bibfnamefont {S.}~\bibnamefont {Sachdev}},
  \bibinfo {author} {\bibfnamefont {M.}~\bibnamefont {Greiner}}, \bibinfo
  {author} {\bibfnamefont {V.}~\bibnamefont {Vuleti{\'c}}}, \ and\ \bibinfo
  {author} {\bibfnamefont {M.~D.}\ \bibnamefont {Lukin}},\ }\href {\doibase
  10.1038/s41586-021-03582-4} {\bibfield  {journal} {\bibinfo  {journal}
  {Nature}\ }\textbf {\bibinfo {volume} {595}},\ \bibinfo {pages} {227}
  (\bibinfo {year} {2021})}\BibitemShut {NoStop}%
\bibitem [{\citenamefont {Iqbal}\ \emph {et~al.}(2024)\citenamefont {Iqbal},
  \citenamefont {Tantivasadakarn}, \citenamefont {Verresen}, \citenamefont
  {Campbell}, \citenamefont {Dreiling}, \citenamefont {Figgatt}, \citenamefont
  {Gaebler}, \citenamefont {Johansen}, \citenamefont {Mills}, \citenamefont
  {Moses}, \citenamefont {Pino}, \citenamefont {Ransford}, \citenamefont
  {Rowe}, \citenamefont {Siegfried}, \citenamefont {Stutz}, \citenamefont
  {Foss-Feig}, \citenamefont {Vishwanath},\ and\ \citenamefont
  {Dreyer}}]{iqbal2024nature}%
  \BibitemOpen
  \bibfield  {author} {\bibinfo {author} {\bibfnamefont {M.}~\bibnamefont
  {Iqbal}}, \bibinfo {author} {\bibfnamefont {N.}~\bibnamefont
  {Tantivasadakarn}}, \bibinfo {author} {\bibfnamefont {R.}~\bibnamefont
  {Verresen}}, \bibinfo {author} {\bibfnamefont {S.~L.}\ \bibnamefont
  {Campbell}}, \bibinfo {author} {\bibfnamefont {J.~M.}\ \bibnamefont
  {Dreiling}}, \bibinfo {author} {\bibfnamefont {C.}~\bibnamefont {Figgatt}},
  \bibinfo {author} {\bibfnamefont {J.~P.}\ \bibnamefont {Gaebler}}, \bibinfo
  {author} {\bibfnamefont {J.}~\bibnamefont {Johansen}}, \bibinfo {author}
  {\bibfnamefont {M.}~\bibnamefont {Mills}}, \bibinfo {author} {\bibfnamefont
  {S.~A.}\ \bibnamefont {Moses}}, \bibinfo {author} {\bibfnamefont {J.~M.}\
  \bibnamefont {Pino}}, \bibinfo {author} {\bibfnamefont {A.}~\bibnamefont
  {Ransford}}, \bibinfo {author} {\bibfnamefont {M.}~\bibnamefont {Rowe}},
  \bibinfo {author} {\bibfnamefont {P.}~\bibnamefont {Siegfried}}, \bibinfo
  {author} {\bibfnamefont {R.~P.}\ \bibnamefont {Stutz}}, \bibinfo {author}
  {\bibfnamefont {M.}~\bibnamefont {Foss-Feig}}, \bibinfo {author}
  {\bibfnamefont {A.}~\bibnamefont {Vishwanath}}, \ and\ \bibinfo {author}
  {\bibfnamefont {H.}~\bibnamefont {Dreyer}},\ }\href {\doibase
  10.1038/s41586-023-06934-4} {\bibfield  {journal} {\bibinfo  {journal}
  {Nature}\ }\textbf {\bibinfo {volume} {626}},\ \bibinfo {pages} {505}
  (\bibinfo {year} {2024})}\BibitemShut {NoStop}%
\bibitem [{\citenamefont {Acharya}\ \emph {et~al.}(2024)\citenamefont
  {Acharya}, \citenamefont {Abanin}, \citenamefont {Aghababaie-Beni},
  \citenamefont {Aleiner}, \citenamefont {Andersen}, \citenamefont {Ansmann},\
  and\ \citenamefont {et~al}}]{acharya2024nature}%
  \BibitemOpen
  \bibfield  {author} {\bibinfo {author} {\bibfnamefont {R.}~\bibnamefont
  {Acharya}}, \bibinfo {author} {\bibfnamefont {D.~A.}\ \bibnamefont {Abanin}},
  \bibinfo {author} {\bibfnamefont {L.}~\bibnamefont {Aghababaie-Beni}},
  \bibinfo {author} {\bibfnamefont {I.}~\bibnamefont {Aleiner}}, \bibinfo
  {author} {\bibfnamefont {T.~I.}\ \bibnamefont {Andersen}}, \bibinfo {author}
  {\bibfnamefont {M.}~\bibnamefont {Ansmann}}, \ and\ \bibinfo {author}
  {\bibnamefont {et~al}},\ }\href {\doibase 10.1038/s41586-024-08449-y}
  {\bibfield  {journal} {\bibinfo  {journal} {Nature}\ } (\bibinfo {year}
  {2024}),\ 10.1038/s41586-024-08449-y}\BibitemShut {NoStop}%
\bibitem [{\citenamefont {Chen}\ \emph {et~al.}(2025)\citenamefont {Chen},
  \citenamefont {Zhu}, \citenamefont {Verresen}, \citenamefont {Seif},
  \citenamefont {B{\"a}umer}, \citenamefont {Layden}, \citenamefont
  {Tantivasadakarn}, \citenamefont {Zhu}, \citenamefont {Sheldon},
  \citenamefont {Vishwanath}, \citenamefont {Trebst},\ and\ \citenamefont
  {Kandala}}]{chen2025np}%
  \BibitemOpen
  \bibfield  {author} {\bibinfo {author} {\bibfnamefont {E.~H.}\ \bibnamefont
  {Chen}}, \bibinfo {author} {\bibfnamefont {G.-Y.}\ \bibnamefont {Zhu}},
  \bibinfo {author} {\bibfnamefont {R.}~\bibnamefont {Verresen}}, \bibinfo
  {author} {\bibfnamefont {A.}~\bibnamefont {Seif}}, \bibinfo {author}
  {\bibfnamefont {E.}~\bibnamefont {B{\"a}umer}}, \bibinfo {author}
  {\bibfnamefont {D.}~\bibnamefont {Layden}}, \bibinfo {author} {\bibfnamefont
  {N.}~\bibnamefont {Tantivasadakarn}}, \bibinfo {author} {\bibfnamefont
  {G.}~\bibnamefont {Zhu}}, \bibinfo {author} {\bibfnamefont {S.}~\bibnamefont
  {Sheldon}}, \bibinfo {author} {\bibfnamefont {A.}~\bibnamefont {Vishwanath}},
  \bibinfo {author} {\bibfnamefont {S.}~\bibnamefont {Trebst}}, \ and\ \bibinfo
  {author} {\bibfnamefont {A.}~\bibnamefont {Kandala}},\ }\href {\doibase
  10.1038/s41567-024-02696-6} {\bibfield  {journal} {\bibinfo  {journal}
  {Nature Physics}\ }\textbf {\bibinfo {volume} {21}},\ \bibinfo {pages} {161}
  (\bibinfo {year} {2025})}\BibitemShut {NoStop}%
\bibitem [{\citenamefont {Verstraete}\ \emph {et~al.}(2004)\citenamefont
  {Verstraete}, \citenamefont {Garc\'{\i}a-Ripoll},\ and\ \citenamefont
  {Cirac}}]{verstrate2004prl}%
  \BibitemOpen
  \bibfield  {author} {\bibinfo {author} {\bibfnamefont {F.}~\bibnamefont
  {Verstraete}}, \bibinfo {author} {\bibfnamefont {J.~J.}\ \bibnamefont
  {Garc\'{\i}a-Ripoll}}, \ and\ \bibinfo {author} {\bibfnamefont {J.~I.}\
  \bibnamefont {Cirac}},\ }\href {\doibase 10.1103/PhysRevLett.93.207204}
  {\bibfield  {journal} {\bibinfo  {journal} {Phys. Rev. Lett.}\ }\textbf
  {\bibinfo {volume} {93}},\ \bibinfo {pages} {207204} (\bibinfo {year}
  {2004})}\BibitemShut {NoStop}%
\bibitem [{\citenamefont {Zwolak}\ and\ \citenamefont
  {Vidal}(2004)}]{zwolak2004prl}%
  \BibitemOpen
  \bibfield  {author} {\bibinfo {author} {\bibfnamefont {M.}~\bibnamefont
  {Zwolak}}\ and\ \bibinfo {author} {\bibfnamefont {G.}~\bibnamefont {Vidal}},\
  }\href {\doibase 10.1103/PhysRevLett.93.207205} {\bibfield  {journal}
  {\bibinfo  {journal} {Phys. Rev. Lett.}\ }\textbf {\bibinfo {volume} {93}},\
  \bibinfo {pages} {207205} (\bibinfo {year} {2004})}\BibitemShut {NoStop}%
\bibitem [{\citenamefont {las Cuevas}\ \emph {et~al.}(2013)\citenamefont {las
  Cuevas}, \citenamefont {Schuch}, \citenamefont {P{\'e}rez-Garc{\'\i}a},\ and\
  \citenamefont {Cirac}}]{cuevas2013njp}%
  \BibitemOpen
  \bibfield  {author} {\bibinfo {author} {\bibfnamefont {G.~D.}\ \bibnamefont
  {las Cuevas}}, \bibinfo {author} {\bibfnamefont {N.}~\bibnamefont {Schuch}},
  \bibinfo {author} {\bibfnamefont {D.}~\bibnamefont {P{\'e}rez-Garc{\'\i}a}},
  \ and\ \bibinfo {author} {\bibfnamefont {J.~I.}\ \bibnamefont {Cirac}},\
  }\href {\doibase 10.1088/1367-2630/15/12/123021} {\bibfield  {journal}
  {\bibinfo  {journal} {New Journal of Physics}\ }\textbf {\bibinfo {volume}
  {15}},\ \bibinfo {pages} {123021} (\bibinfo {year} {2013})}\BibitemShut
  {NoStop}%
\bibitem [{\citenamefont {Pollmann}\ \emph {et~al.}(2010)\citenamefont
  {Pollmann}, \citenamefont {Turner}, \citenamefont {Berg},\ and\ \citenamefont
  {Oshikawa}}]{pollmann2010prb}%
  \BibitemOpen
  \bibfield  {author} {\bibinfo {author} {\bibfnamefont {F.}~\bibnamefont
  {Pollmann}}, \bibinfo {author} {\bibfnamefont {A.~M.}\ \bibnamefont
  {Turner}}, \bibinfo {author} {\bibfnamefont {E.}~\bibnamefont {Berg}}, \ and\
  \bibinfo {author} {\bibfnamefont {M.}~\bibnamefont {Oshikawa}},\ }\href
  {\doibase 10.1103/PhysRevB.81.064439} {\bibfield  {journal} {\bibinfo
  {journal} {Phys. Rev. B}\ }\textbf {\bibinfo {volume} {81}},\ \bibinfo
  {pages} {064439} (\bibinfo {year} {2010})}\BibitemShut {NoStop}%
\bibitem [{\citenamefont {Chen}\ \emph {et~al.}(2011)\citenamefont {Chen},
  \citenamefont {Gu},\ and\ \citenamefont {Wen}}]{chen2011prb}%
  \BibitemOpen
  \bibfield  {author} {\bibinfo {author} {\bibfnamefont {X.}~\bibnamefont
  {Chen}}, \bibinfo {author} {\bibfnamefont {Z.-C.}\ \bibnamefont {Gu}}, \ and\
  \bibinfo {author} {\bibfnamefont {X.-G.}\ \bibnamefont {Wen}},\ }\href
  {\doibase 10.1103/PhysRevB.83.035107} {\bibfield  {journal} {\bibinfo
  {journal} {Phys. Rev. B}\ }\textbf {\bibinfo {volume} {83}},\ \bibinfo
  {pages} {035107} (\bibinfo {year} {2011})}\BibitemShut {NoStop}%
\bibitem [{\citenamefont {Schuch}\ \emph {et~al.}(2011)\citenamefont {Schuch},
  \citenamefont {P\'erez-Garc\'{\i}a},\ and\ \citenamefont
  {Cirac}}]{schuch2011prb}%
  \BibitemOpen
  \bibfield  {author} {\bibinfo {author} {\bibfnamefont {N.}~\bibnamefont
  {Schuch}}, \bibinfo {author} {\bibfnamefont {D.}~\bibnamefont
  {P\'erez-Garc\'{\i}a}}, \ and\ \bibinfo {author} {\bibfnamefont
  {I.}~\bibnamefont {Cirac}},\ }\href {\doibase 10.1103/PhysRevB.84.165139}
  {\bibfield  {journal} {\bibinfo  {journal} {Phys. Rev. B}\ }\textbf {\bibinfo
  {volume} {84}},\ \bibinfo {pages} {165139} (\bibinfo {year}
  {2011})}\BibitemShut {NoStop}%
\bibitem [{\citenamefont {Senthil}(2015)}]{senthil2015arcmp}%
  \BibitemOpen
  \bibfield  {author} {\bibinfo {author} {\bibfnamefont {T.}~\bibnamefont
  {Senthil}},\ }\href {\doibase
  https://doi.org/10.1146/annurev-conmatphys-031214-014740} {\bibfield
  {journal} {\bibinfo  {journal} {Annual Review of Condensed Matter Physics}\
  }\textbf {\bibinfo {volume} {6}},\ \bibinfo {pages} {299} (\bibinfo {year}
  {2015})}\BibitemShut {NoStop}%
\bibitem [{\citenamefont {Zaletel}\ \emph {et~al.}(2013)\citenamefont
  {Zaletel}, \citenamefont {Mong},\ and\ \citenamefont
  {Pollmann}}]{zaletel2013prl}%
  \BibitemOpen
  \bibfield  {author} {\bibinfo {author} {\bibfnamefont {M.~P.}\ \bibnamefont
  {Zaletel}}, \bibinfo {author} {\bibfnamefont {R.~S.~K.}\ \bibnamefont
  {Mong}}, \ and\ \bibinfo {author} {\bibfnamefont {F.}~\bibnamefont
  {Pollmann}},\ }\href {\doibase 10.1103/PhysRevLett.110.236801} {\bibfield
  {journal} {\bibinfo  {journal} {Phys. Rev. Lett.}\ }\textbf {\bibinfo
  {volume} {110}},\ \bibinfo {pages} {236801} (\bibinfo {year}
  {2013})}\BibitemShut {NoStop}%
\bibitem [{\citenamefont {Zaletel}\ \emph {et~al.}(2014)\citenamefont
  {Zaletel}, \citenamefont {Mong},\ and\ \citenamefont
  {Pollmann}}]{zaletel2014njp}%
  \BibitemOpen
  \bibfield  {author} {\bibinfo {author} {\bibfnamefont {M.~P.}\ \bibnamefont
  {Zaletel}}, \bibinfo {author} {\bibfnamefont {R.~S.~K.}\ \bibnamefont
  {Mong}}, \ and\ \bibinfo {author} {\bibfnamefont {F.}~\bibnamefont
  {Pollmann}},\ }\href {\doibase 10.1088/1742-5468/2014/10/P10007} {\bibfield
  {journal} {\bibinfo  {journal} {Journal of Statistical Mechanics: Theory and
  Experiment}\ }\textbf {\bibinfo {volume} {2014}},\ \bibinfo {pages} {P10007}
  (\bibinfo {year} {2014})}\BibitemShut {NoStop}%
\bibitem [{\citenamefont {Zaletel}(2014)}]{zaletel2014prb}%
  \BibitemOpen
  \bibfield  {author} {\bibinfo {author} {\bibfnamefont {M.~P.}\ \bibnamefont
  {Zaletel}},\ }\href {\doibase 10.1103/PhysRevB.90.235113} {\bibfield
  {journal} {\bibinfo  {journal} {Phys. Rev. B}\ }\textbf {\bibinfo {volume}
  {90}},\ \bibinfo {pages} {235113} (\bibinfo {year} {2014})}\BibitemShut
  {NoStop}%
\bibitem [{\citenamefont {Cirac}\ \emph {et~al.}(2021)\citenamefont {Cirac},
  \citenamefont {P\'erez-Garc\'{\i}a}, \citenamefont {Schuch},\ and\
  \citenamefont {Verstraete}}]{cirac2021rmp}%
  \BibitemOpen
  \bibfield  {author} {\bibinfo {author} {\bibfnamefont {J.~I.}\ \bibnamefont
  {Cirac}}, \bibinfo {author} {\bibfnamefont {D.}~\bibnamefont
  {P\'erez-Garc\'{\i}a}}, \bibinfo {author} {\bibfnamefont {N.}~\bibnamefont
  {Schuch}}, \ and\ \bibinfo {author} {\bibfnamefont {F.}~\bibnamefont
  {Verstraete}},\ }\href {\doibase 10.1103/RevModPhys.93.045003} {\bibfield
  {journal} {\bibinfo  {journal} {Rev. Mod. Phys.}\ }\textbf {\bibinfo {volume}
  {93}},\ \bibinfo {pages} {045003} (\bibinfo {year} {2021})}\BibitemShut
  {NoStop}%
\bibitem [{Note1()}]{Note1}%
  \BibitemOpen
  \bibinfo {note} {Here the requirement of a well-defined linear on-site
  representation is to ensure the stability of SPTs~\cite
  {pollmann2017oxford,anfuso2007fragility}.}\BibitemShut {Stop}%
\bibitem [{\citenamefont {P\'erez-Garc\'{\i}a}\ \emph
  {et~al.}(2008)\citenamefont {P\'erez-Garc\'{\i}a}, \citenamefont {Wolf},
  \citenamefont {Sanz}, \citenamefont {Verstraete},\ and\ \citenamefont
  {Cirac}}]{garcia2008prl}%
  \BibitemOpen
  \bibfield  {author} {\bibinfo {author} {\bibfnamefont {D.}~\bibnamefont
  {P\'erez-Garc\'{\i}a}}, \bibinfo {author} {\bibfnamefont {M.~M.}\
  \bibnamefont {Wolf}}, \bibinfo {author} {\bibfnamefont {M.}~\bibnamefont
  {Sanz}}, \bibinfo {author} {\bibfnamefont {F.}~\bibnamefont {Verstraete}}, \
  and\ \bibinfo {author} {\bibfnamefont {J.~I.}\ \bibnamefont {Cirac}},\ }\href
  {\doibase 10.1103/PhysRevLett.100.167202} {\bibfield  {journal} {\bibinfo
  {journal} {Phys. Rev. Lett.}\ }\textbf {\bibinfo {volume} {100}},\ \bibinfo
  {pages} {167202} (\bibinfo {year} {2008})}\BibitemShut {NoStop}%
\bibitem [{\citenamefont {Chen}\ \emph {et~al.}(2013)\citenamefont {Chen},
  \citenamefont {Gu}, \citenamefont {Liu},\ and\ \citenamefont
  {Wen}}]{chen2013prb}%
  \BibitemOpen
  \bibfield  {author} {\bibinfo {author} {\bibfnamefont {X.}~\bibnamefont
  {Chen}}, \bibinfo {author} {\bibfnamefont {Z.-C.}\ \bibnamefont {Gu}},
  \bibinfo {author} {\bibfnamefont {Z.-X.}\ \bibnamefont {Liu}}, \ and\
  \bibinfo {author} {\bibfnamefont {X.-G.}\ \bibnamefont {Wen}},\ }\href
  {\doibase 10.1103/PhysRevB.87.155114} {\bibfield  {journal} {\bibinfo
  {journal} {Phys. Rev. B}\ }\textbf {\bibinfo {volume} {87}},\ \bibinfo
  {pages} {155114} (\bibinfo {year} {2013})}\BibitemShut {NoStop}%
\bibitem [{sup()}]{suppupdated}%
  \BibitemOpen
  \href@noop {} {}\bibinfo {note} {The Supplemental Material provides: (i)
  implementation details of weak symmetry flux insertion; (ii) a proof of the
  selection rule for string order parameters; (iii) a review of average
  symmetry-protected topological (ASPT) phases in one dimension, via quantized
  response; and (iv) an identity for the conservation of topological charge
  between physical and ancillary systems.}\BibitemShut {Stop}%
\bibitem [{\citenamefont {den Nijs}\ and\ \citenamefont
  {Rommelse}(1989)}]{nijs1989prb}%
  \BibitemOpen
  \bibfield  {author} {\bibinfo {author} {\bibfnamefont {M.}~\bibnamefont {den
  Nijs}}\ and\ \bibinfo {author} {\bibfnamefont {K.}~\bibnamefont {Rommelse}},\
  }\href {\doibase 10.1103/PhysRevB.40.4709} {\bibfield  {journal} {\bibinfo
  {journal} {Phys. Rev. B}\ }\textbf {\bibinfo {volume} {40}},\ \bibinfo
  {pages} {4709} (\bibinfo {year} {1989})}\BibitemShut {NoStop}%
\bibitem [{\citenamefont {Kennedy}\ and\ \citenamefont
  {Tasaki}(1992)}]{kennedy1992prb}%
  \BibitemOpen
  \bibfield  {author} {\bibinfo {author} {\bibfnamefont {T.}~\bibnamefont
  {Kennedy}}\ and\ \bibinfo {author} {\bibfnamefont {H.}~\bibnamefont
  {Tasaki}},\ }\href {\doibase 10.1103/PhysRevB.45.304} {\bibfield  {journal}
  {\bibinfo  {journal} {Phys. Rev. B}\ }\textbf {\bibinfo {volume} {45}},\
  \bibinfo {pages} {304} (\bibinfo {year} {1992})}\BibitemShut {NoStop}%
\bibitem [{\citenamefont {Pollmann}\ and\ \citenamefont
  {Turner}(2012)}]{pollmann2012prb}%
  \BibitemOpen
  \bibfield  {author} {\bibinfo {author} {\bibfnamefont {F.}~\bibnamefont
  {Pollmann}}\ and\ \bibinfo {author} {\bibfnamefont {A.~M.}\ \bibnamefont
  {Turner}},\ }\href {\doibase 10.1103/PhysRevB.86.125441} {\bibfield
  {journal} {\bibinfo  {journal} {Phys. Rev. B}\ }\textbf {\bibinfo {volume}
  {86}},\ \bibinfo {pages} {125441} (\bibinfo {year} {2012})}\BibitemShut
  {NoStop}%
\bibitem [{\citenamefont {Fredenhagen}\ and\ \citenamefont
  {Marcu}(1983)}]{fredenhagen1983cmp}%
  \BibitemOpen
  \bibfield  {author} {\bibinfo {author} {\bibfnamefont {K.}~\bibnamefont
  {Fredenhagen}}\ and\ \bibinfo {author} {\bibfnamefont {M.}~\bibnamefont
  {Marcu}},\ }\href {\doibase 10.1007/BF01206315} {\bibfield  {journal}
  {\bibinfo  {journal} {Communications in Mathematical Physics}\ }\textbf
  {\bibinfo {volume} {92}},\ \bibinfo {pages} {81} (\bibinfo {year}
  {1983})}\BibitemShut {NoStop}%
\bibitem [{\citenamefont {Marcu}(1986)}]{marcu1986springer}%
  \BibitemOpen
  \bibfield  {author} {\bibinfo {author} {\bibfnamefont {M.}~\bibnamefont
  {Marcu}},\ }\enquote {\bibinfo {title} {(uses of) an order parameter for
  lattice gauge theories with matter fields},}\ in\ \href {\doibase
  10.1007/978-1-4613-2231-3_25} {\emph {\bibinfo {booktitle} {Lattice Gauge
  Theory: A Challenge in Large-Scale Computing}}},\ \bibinfo {editor} {edited
  by\ \bibinfo {editor} {\bibfnamefont {B.}~\bibnamefont {Bunk}}, \bibinfo
  {editor} {\bibfnamefont {K.~H.}\ \bibnamefont {M{\"u}tter}}, \ and\ \bibinfo
  {editor} {\bibfnamefont {K.}~\bibnamefont {Schilling}}}\ (\bibinfo
  {publisher} {Springer US},\ \bibinfo {address} {Boston, MA},\ \bibinfo {year}
  {1986})\ pp.\ \bibinfo {pages} {267--278}\BibitemShut {NoStop}%
\bibitem [{\citenamefont {Xu}\ \emph {et~al.}(2024)\citenamefont {Xu},
  \citenamefont {Pollmann},\ and\ \citenamefont {Knap}}]{xu2024arxiv}%
  \BibitemOpen
  \bibfield  {author} {\bibinfo {author} {\bibfnamefont {W.-T.}\ \bibnamefont
  {Xu}}, \bibinfo {author} {\bibfnamefont {F.}~\bibnamefont {Pollmann}}, \ and\
  \bibinfo {author} {\bibfnamefont {M.}~\bibnamefont {Knap}},\ }\href@noop {}
  {\enquote {\bibinfo {title} {Critical behavior of fredenhagen-marcu string
  order parameters at topological phase transitions with emergent higher-form
  symmetries},}\ } (\bibinfo {year} {2024}),\ \Eprint
  {http://arxiv.org/abs/2402.00127} {arXiv:2402.00127 [cond-mat.str-el]}
  \BibitemShut {NoStop}%
\bibitem [{\citenamefont {Jiang}\ \emph {et~al.}(2023)\citenamefont {Jiang},
  \citenamefont {Chen}, \citenamefont {Liu}, \citenamefont {Rong},
  \citenamefont {Assaad}, \citenamefont {Cheng}, \citenamefont {Sun},\ and\
  \citenamefont {Meng}}]{jiang2023scipost}%
  \BibitemOpen
  \bibfield  {author} {\bibinfo {author} {\bibfnamefont {W.}~\bibnamefont
  {Jiang}}, \bibinfo {author} {\bibfnamefont {B.-B.}\ \bibnamefont {Chen}},
  \bibinfo {author} {\bibfnamefont {Z.~H.}\ \bibnamefont {Liu}}, \bibinfo
  {author} {\bibfnamefont {J.}~\bibnamefont {Rong}}, \bibinfo {author}
  {\bibfnamefont {F.~F.}\ \bibnamefont {Assaad}}, \bibinfo {author}
  {\bibfnamefont {M.}~\bibnamefont {Cheng}}, \bibinfo {author} {\bibfnamefont
  {K.}~\bibnamefont {Sun}}, \ and\ \bibinfo {author} {\bibfnamefont {Z.~Y.}\
  \bibnamefont {Meng}},\ }\href {\doibase 10.21468/SciPostPhys.15.3.082}
  {\bibfield  {journal} {\bibinfo  {journal} {SciPost Phys.}\ }\textbf
  {\bibinfo {volume} {15}},\ \bibinfo {pages} {082} (\bibinfo {year}
  {2023})}\BibitemShut {NoStop}%
\bibitem [{\citenamefont {Wang}\ \emph {et~al.}(2024)\citenamefont {Wang},
  \citenamefont {Zhou}, \citenamefont {Zhou},\ and\ \citenamefont
  {Zhang}}]{wang2024prl}%
  \BibitemOpen
  \bibfield  {author} {\bibinfo {author} {\bibfnamefont {C.-Y.}\ \bibnamefont
  {Wang}}, \bibinfo {author} {\bibfnamefont {T.-G.}\ \bibnamefont {Zhou}},
  \bibinfo {author} {\bibfnamefont {Y.-N.}\ \bibnamefont {Zhou}}, \ and\
  \bibinfo {author} {\bibfnamefont {P.}~\bibnamefont {Zhang}},\ }\href
  {\doibase 10.1103/PhysRevLett.133.083402} {\bibfield  {journal} {\bibinfo
  {journal} {Phys. Rev. Lett.}\ }\textbf {\bibinfo {volume} {133}},\ \bibinfo
  {pages} {083402} (\bibinfo {year} {2024})}\BibitemShut {NoStop}%
\bibitem [{\citenamefont {Zang}\ \emph {et~al.}(2024)\citenamefont {Zang},
  \citenamefont {Gu},\ and\ \citenamefont {Jiang}}]{zang2024prl}%
  \BibitemOpen
  \bibfield  {author} {\bibinfo {author} {\bibfnamefont {Y.}~\bibnamefont
  {Zang}}, \bibinfo {author} {\bibfnamefont {Y.}~\bibnamefont {Gu}}, \ and\
  \bibinfo {author} {\bibfnamefont {S.}~\bibnamefont {Jiang}},\ }\href
  {\doibase 10.1103/PhysRevLett.133.106503} {\bibfield  {journal} {\bibinfo
  {journal} {Phys. Rev. Lett.}\ }\textbf {\bibinfo {volume} {133}},\ \bibinfo
  {pages} {106503} (\bibinfo {year} {2024})}\BibitemShut {NoStop}%
\bibitem [{\citenamefont {Mao}\ \emph {et~al.}(2024{\natexlab{b}})\citenamefont
  {Mao}, \citenamefont {Zhai},\ and\ \citenamefont {Yang}}]{mao2024arxiv}%
  \BibitemOpen
  \bibfield  {author} {\bibinfo {author} {\bibfnamefont {L.}~\bibnamefont
  {Mao}}, \bibinfo {author} {\bibfnamefont {H.}~\bibnamefont {Zhai}}, \ and\
  \bibinfo {author} {\bibfnamefont {F.}~\bibnamefont {Yang}},\ }\href@noop {}
  {\enquote {\bibinfo {title} {Probing topology of gaussian mixed states by the
  full counting statistics},}\ } (\bibinfo {year} {2024}{\natexlab{b}}),\
  \Eprint {http://arxiv.org/abs/2402.15964} {arXiv:2402.15964
  [cond-mat.mes-hall]} \BibitemShut {NoStop}%
\bibitem [{\citenamefont {Pollmann}(2017)}]{pollmann2017oxford}%
  \BibitemOpen
  \bibfield  {author} {\bibinfo {author} {\bibfnamefont {F.}~\bibnamefont
  {Pollmann}},\ }in\ \href@noop {} {\emph {\bibinfo {booktitle} {Topological
  Aspects of Condensed Matter Physics: Lecture Notes of the Les Houches Summer
  School 2014}}},\ Vol.\ \bibinfo {volume} {103}\ (\bibinfo  {publisher}
  {Oxford University Press},\ \bibinfo {address} {Oxford},\ \bibinfo {year}
  {2017})\BibitemShut {NoStop}%
\bibitem [{\citenamefont {Nielsen}\ and\ \citenamefont
  {Chuang}(2012)}]{nielsen2010cambridge}%
  \BibitemOpen
  \bibfield  {author} {\bibinfo {author} {\bibfnamefont {M.~A.}\ \bibnamefont
  {Nielsen}}\ and\ \bibinfo {author} {\bibfnamefont {I.~L.}\ \bibnamefont
  {Chuang}},\ }\href@noop {} {\emph {\bibinfo {title} {Quantum Computation and
  Quantum Information}}}\ (\bibinfo  {publisher} {Cambridge University Press},\
  \bibinfo {address} {Cambridge, England},\ \bibinfo {year} {2012})\BibitemShut
  {NoStop}%
\bibitem [{fn_()}]{fn_fig2}%
  \BibitemOpen
  \href@noop {} {}\bibinfo {note} {The decay exponent plot range is $p\in
  [0.3,\ 0.7]$, selected to avoid numerical instability from accidental zeros
  in $\mathcal{S}_y$ at $p=\frac{1}{4}$. Analytical results are provided in the
  Appendix B for large string lengths.}\BibitemShut {Stop}%
\bibitem [{\citenamefont {Ruiz-de Alarc{\'o}n}\ \emph
  {et~al.}(2024)\citenamefont {Ruiz-de Alarc{\'o}n}, \citenamefont
  {Garre-Rubio}, \citenamefont {Moln{\'a}r},\ and\ \citenamefont
  {P{\'e}rez-Garc{\'\i}a}}]{ruiz2024lmp}%
  \BibitemOpen
  \bibfield  {author} {\bibinfo {author} {\bibfnamefont {A.}~\bibnamefont
  {Ruiz-de Alarc{\'o}n}}, \bibinfo {author} {\bibfnamefont {J.}~\bibnamefont
  {Garre-Rubio}}, \bibinfo {author} {\bibfnamefont {A.}~\bibnamefont
  {Moln{\'a}r}}, \ and\ \bibinfo {author} {\bibfnamefont {D.}~\bibnamefont
  {P{\'e}rez-Garc{\'\i}a}},\ }\href {\doibase 10.1007/s11005-024-01778-z}
  {\bibfield  {journal} {\bibinfo  {journal} {Letters in Mathematical Physics}\
  }\textbf {\bibinfo {volume} {114}},\ \bibinfo {pages} {43} (\bibinfo {year}
  {2024})}\BibitemShut {NoStop}%
\bibitem [{\citenamefont {Hasan}\ and\ \citenamefont
  {Kane}(2010)}]{hassan2010rmp}%
  \BibitemOpen
  \bibfield  {author} {\bibinfo {author} {\bibfnamefont {M.~Z.}\ \bibnamefont
  {Hasan}}\ and\ \bibinfo {author} {\bibfnamefont {C.~L.}\ \bibnamefont
  {Kane}},\ }\href {\doibase 10.1103/RevModPhys.82.3045} {\bibfield  {journal}
  {\bibinfo  {journal} {Rev. Mod. Phys.}\ }\textbf {\bibinfo {volume} {82}},\
  \bibinfo {pages} {3045} (\bibinfo {year} {2010})}\BibitemShut {NoStop}%
\bibitem [{\citenamefont {Qi}\ and\ \citenamefont {Zhang}(2011)}]{qi2011rmp}%
  \BibitemOpen
  \bibfield  {author} {\bibinfo {author} {\bibfnamefont {X.-L.}\ \bibnamefont
  {Qi}}\ and\ \bibinfo {author} {\bibfnamefont {S.-C.}\ \bibnamefont {Zhang}},\
  }\href {\doibase 10.1103/RevModPhys.83.1057} {\bibfield  {journal} {\bibinfo
  {journal} {Rev. Mod. Phys.}\ }\textbf {\bibinfo {volume} {83}},\ \bibinfo
  {pages} {1057} (\bibinfo {year} {2011})}\BibitemShut {NoStop}%
\bibitem [{\citenamefont {Endres}\ \emph {et~al.}(2011)\citenamefont {Endres},
  \citenamefont {Cheneau}, \citenamefont {Fukuhara}, \citenamefont
  {Weitenberg}, \citenamefont {Schau{\ss}}, \citenamefont {Gross},
  \citenamefont {Mazza}, \citenamefont {Ba{\~n}uls}, \citenamefont {Pollet},
  \citenamefont {Bloch},\ and\ \citenamefont {Kuhr}}]{endres2011science}%
  \BibitemOpen
  \bibfield  {author} {\bibinfo {author} {\bibfnamefont {M.}~\bibnamefont
  {Endres}}, \bibinfo {author} {\bibfnamefont {M.}~\bibnamefont {Cheneau}},
  \bibinfo {author} {\bibfnamefont {T.}~\bibnamefont {Fukuhara}}, \bibinfo
  {author} {\bibfnamefont {C.}~\bibnamefont {Weitenberg}}, \bibinfo {author}
  {\bibfnamefont {P.}~\bibnamefont {Schau{\ss}}}, \bibinfo {author}
  {\bibfnamefont {C.}~\bibnamefont {Gross}}, \bibinfo {author} {\bibfnamefont
  {L.}~\bibnamefont {Mazza}}, \bibinfo {author} {\bibfnamefont {M.~C.}\
  \bibnamefont {Ba{\~n}uls}}, \bibinfo {author} {\bibfnamefont
  {L.}~\bibnamefont {Pollet}}, \bibinfo {author} {\bibfnamefont
  {I.}~\bibnamefont {Bloch}}, \ and\ \bibinfo {author} {\bibfnamefont
  {S.}~\bibnamefont {Kuhr}},\ }\href {\doibase 10.1126/science.1209284}
  {\bibfield  {journal} {\bibinfo  {journal} {Science}\ }\textbf {\bibinfo
  {volume} {334}},\ \bibinfo {pages} {200} (\bibinfo {year}
  {2011})}\BibitemShut {NoStop}%
\bibitem [{\citenamefont {Karch}\ \emph {et~al.}(2025)\citenamefont {Karch},
  \citenamefont {Bandyopadhyay}, \citenamefont {Sun}, \citenamefont {Impertro},
  \citenamefont {Huh}, \citenamefont {Rodríguez}, \citenamefont {Wienand},
  \citenamefont {Ketterle}, \citenamefont {Heyl}, \citenamefont {Polkovnikov},
  \citenamefont {Bloch},\ and\ \citenamefont {Aidelsburger}}]{karch2025arxiv}%
  \BibitemOpen
  \bibfield  {author} {\bibinfo {author} {\bibfnamefont {S.}~\bibnamefont
  {Karch}}, \bibinfo {author} {\bibfnamefont {S.}~\bibnamefont
  {Bandyopadhyay}}, \bibinfo {author} {\bibfnamefont {Z.-H.}\ \bibnamefont
  {Sun}}, \bibinfo {author} {\bibfnamefont {A.}~\bibnamefont {Impertro}},
  \bibinfo {author} {\bibfnamefont {S.}~\bibnamefont {Huh}}, \bibinfo {author}
  {\bibfnamefont {I.~P.}\ \bibnamefont {Rodríguez}}, \bibinfo {author}
  {\bibfnamefont {J.~F.}\ \bibnamefont {Wienand}}, \bibinfo {author}
  {\bibfnamefont {W.}~\bibnamefont {Ketterle}}, \bibinfo {author}
  {\bibfnamefont {M.}~\bibnamefont {Heyl}}, \bibinfo {author} {\bibfnamefont
  {A.}~\bibnamefont {Polkovnikov}}, \bibinfo {author} {\bibfnamefont
  {I.}~\bibnamefont {Bloch}}, \ and\ \bibinfo {author} {\bibfnamefont
  {M.}~\bibnamefont {Aidelsburger}},\ }\href@noop {} {\enquote {\bibinfo
  {title} {Probing quantum many-body dynamics using subsystem loschmidt
  echos},}\ } (\bibinfo {year} {2025}),\ \Eprint
  {http://arxiv.org/abs/2501.16995} {arXiv:2501.16995 [cond-mat.quant-gas]}
  \BibitemShut {NoStop}%
\bibitem [{\citenamefont {Fan}\ \emph {et~al.}(2024)\citenamefont {Fan},
  \citenamefont {Bao}, \citenamefont {Altman},\ and\ \citenamefont
  {Vishwanath}}]{fan2024prxQ}%
  \BibitemOpen
  \bibfield  {author} {\bibinfo {author} {\bibfnamefont {R.}~\bibnamefont
  {Fan}}, \bibinfo {author} {\bibfnamefont {Y.}~\bibnamefont {Bao}}, \bibinfo
  {author} {\bibfnamefont {E.}~\bibnamefont {Altman}}, \ and\ \bibinfo {author}
  {\bibfnamefont {A.}~\bibnamefont {Vishwanath}},\ }\href {\doibase
  10.1103/PRXQuantum.5.020343} {\bibfield  {journal} {\bibinfo  {journal} {PRX
  Quantum}\ }\textbf {\bibinfo {volume} {5}},\ \bibinfo {pages} {020343}
  (\bibinfo {year} {2024})}\BibitemShut {NoStop}%
\bibitem [{\citenamefont {Huang}\ \emph {et~al.}(2024)\citenamefont {Huang},
  \citenamefont {Colmenarez}, \citenamefont {Müller},\ and\ \citenamefont
  {Diehl}}]{huang2024CI}%
  \BibitemOpen
  \bibfield  {author} {\bibinfo {author} {\bibfnamefont {Z.-M.}\ \bibnamefont
  {Huang}}, \bibinfo {author} {\bibfnamefont {L.}~\bibnamefont {Colmenarez}},
  \bibinfo {author} {\bibfnamefont {M.}~\bibnamefont {Müller}}, \ and\
  \bibinfo {author} {\bibfnamefont {S.}~\bibnamefont {Diehl}},\ }\href@noop {}
  {\enquote {\bibinfo {title} {Coherent information as a mixed-state
  topological order parameter of fermions},}\ } (\bibinfo {year} {2024}),\
  \Eprint {http://arxiv.org/abs/2412.12279} {arXiv:2412.12279 [quant-ph]}
  \BibitemShut {NoStop}%
\bibitem [{\citenamefont {Lee}\ \emph {et~al.}(2023)\citenamefont {Lee},
  \citenamefont {Jian},\ and\ \citenamefont {Xu}}]{lee2023prxq}%
  \BibitemOpen
  \bibfield  {author} {\bibinfo {author} {\bibfnamefont {J.~Y.}\ \bibnamefont
  {Lee}}, \bibinfo {author} {\bibfnamefont {C.-M.}\ \bibnamefont {Jian}}, \
  and\ \bibinfo {author} {\bibfnamefont {C.}~\bibnamefont {Xu}},\ }\href
  {\doibase 10.1103/PRXQuantum.4.030317} {\bibfield  {journal} {\bibinfo
  {journal} {PRX Quantum}\ }\textbf {\bibinfo {volume} {4}},\ \bibinfo {pages}
  {030317} (\bibinfo {year} {2023})}\BibitemShut {NoStop}%
\bibitem [{\citenamefont {Behrends}\ and\ \citenamefont
  {Béri}(2024)}]{behrends2024arxiv}%
  \BibitemOpen
  \bibfield  {author} {\bibinfo {author} {\bibfnamefont {J.}~\bibnamefont
  {Behrends}}\ and\ \bibinfo {author} {\bibfnamefont {B.}~\bibnamefont
  {Béri}},\ }\href@noop {} {\enquote {\bibinfo {title} {The surface code under
  generic $x$-error channels: Statistical mechanics, error thresholds, and
  errorfield double phenomenology},}\ } (\bibinfo {year} {2024}),\ \Eprint
  {http://arxiv.org/abs/2412.21055} {arXiv:2412.21055 [quant-ph]} \BibitemShut
  {NoStop}%
\bibitem [{\citenamefont {Anfuso}\ and\ \citenamefont
  {Rosch}(2007)}]{anfuso2007fragility}%
  \BibitemOpen
  \bibfield  {author} {\bibinfo {author} {\bibfnamefont {F.}~\bibnamefont
  {Anfuso}}\ and\ \bibinfo {author} {\bibfnamefont {A.}~\bibnamefont {Rosch}},\
  }\href {\doibase 10.1103/PhysRevB.76.085124} {\bibfield  {journal} {\bibinfo
  {journal} {Phys. Rev. B}\ }\textbf {\bibinfo {volume} {76}},\ \bibinfo
  {pages} {085124} (\bibinfo {year} {2007})}\BibitemShut {NoStop}%
\end{thebibliography}
\end{document}